\documentclass[%
superscriptaddress,
a4paper,cite,
twocolumn,
 amsmath,amssymb,
 aps,
prb,
]{revtex4-2}

\usepackage{xcolor}
\usepackage{xcolor}
\usepackage{cancel}
\usepackage[T1]{fontenc}
\usepackage{siunitx}
\usepackage{graphicx}
\usepackage{dcolumn}
\usepackage{bm}

\newcommand{\kb}[0]{\mathbf{k}}

\newcommand{\be}{\begin{equation}}
\newcommand{\ee}{\end{equation}}
\newcommand{\bea}{\begin{eqnarray}}
\newcommand{\eea}{\end{eqnarray}}
\newcommand{\bean}{\begin{eqnarray*}}
\newcommand{\eean}{\end{eqnarray*}}

\usepackage[capitalise]{cleveref}
\usepackage[normalem]{ulem}

\begin{document}

\preprint{APS/123-QED}

\title{GW effects on the  topology of  type-II  Dirac cones in  NiTe$_2$, PtSe$_2$ and PtTe$_2$
}

\author{Franz Fischer}
\affiliation{Physical Chemistry and Physics Departments, University of Hamburg, 
Hamburg 22761, Germany }
\affiliation{Max Planck Institute for the Structure and Dynamics of Matter, 22761 Hamburg, Germany}
\author{Abderrezak Torche}
\affiliation{Physical Chemistry and Physics Departments, University of Hamburg, 
Hamburg 22761, Germany }
\author{Marta Prada}
\affiliation{I Institute for Theoretical Physics, University of Hamburg, 
Hamburg 22761, Germany }
\affiliation{Physical Chemistry and Physics Departments, University of Hamburg, 
Hamburg 22761, Germany }
\email{mprada@physnet.uni-hamburg.de}
\author{Gabriel Bester}
\affiliation{Physical Chemistry and Physics Departments, University of Hamburg, 
Hamburg 22761, Germany }
\affiliation{The Hamburg Centre for Ultrafast Imaging, Luruper Chaussee 149, 22761 Hamburg, Germany.}

\begin{abstract}
Many-body correlations are known to be responsible for a broad range of fascinating physical phenomena, introducing corrections that appear elusive at
the single-particle level. 
An example of this is the Lifshitz transition that occurs as the Fermi surface topology changes when {\it e.g.} Coulomb interaction effects break into the picture. In particular,  the Fermi velocity renormalization can lead a type-II Weyl semimetal at mean-field level to become a trivial or a type-I Dirac material when correlations are accounted for, which is far from being obvious. 
In this work we scrutinize the band structure of NiTe$_2$, a material that features a type-II Dirac point near the Fermi level within the mean-field approach. 
Including GW-level correlations, our findings showcase anisotropic corrections on the Dirac carrier velocity exceeding $100 \, \%$ 
enhancements, underscoring the nuanced influence of electronic interactions in the band structure.  
We also consider type-II Dirac crossings in  PtSe$_2$ and  PtTe$_2$ and observe that including many-body effects via GW the band topology changes, featuring trivial topology and type-I Dirac crossings, respectively. Our findings highlights the necessity to evaluate the many-body effects on non-trivial bands, 
contributing essential insights to the broader exploration of many-body correlation effects in type-II Dirac points of condensed-matter systems.
\end{abstract}

\maketitle

\section{Introduction}
\label{sec:introduction}

The rise of graphene in 2004 \cite{novoselov2004electric,geim2010rise} and the subsequent discovery of topological insulators \cite{hasan2010colloquium} have brought 
the study of Dirac materials to the forefront of contemporary research  \cite{giustino20212021,wehling2014dirac}. 
The insulating character of the bulk with intriguing conducting edges led to a whole new paradigm of novel states of matter that are mostly probed in 
transport, magnetic and thermodynamic experiments \cite{wehling2014dirac}. Recently, a new class of materials has been found within this family, namely the so-called 
type-II Dirac materials \cite{soluyanov2015type}. Whereas in type-I structures the Fermi surface can be reduced to a point,  
type-II Dirac fermions show an over-tilted conic dispersion, such that the Fermi surface (FS) forms an open arc 
where electron and hole pockets touch when plotted in two dimensions (see Fig. \ref{fig:figPic}). This Fermi arc indicate the presence of surface states in the material with  exceptional properties.
The nontrivial topology involves the presence of topologically protected surface states and bulk band crossings, 
giving rise to interesting phenomena that can be exploited for novel electronic and spintronic devices: 
The  anisotropic chiral anomaly \cite{udagawa2016field}, unusual magnetoresponse with Landau level collapse \cite{yu2016predicted} and novel 
quantum oscillations \cite{o2016magnetic} are a few examples. 
From a high-energy physics perspective, as type-II Dirac electrons violate Lorentz invariance, it has been suggested that the low-energy excitations of these 
Dirac materials present a parallelism with those of quantum field theories of curved spaces \cite{volovik2016black}.


Type-II Dirac materials are nowadays a fast-growing family that includes 
many transition metal dichalcogenides (TMDs) (WTe$_2$ , NiTe$_2$ , PtTe$_2$ , PtSe$_2$) \cite{ soluyanov2015type,huang2016type,yan2017lorentz}, transition metal  icosagenides: MA$_3$   
(M=V, Nb, Ta and A=Al, Ga, In) \cite{chang2017type} and some Full-Heusler compounds  XMg$_2$Ag (with X=Pr, Nd, Sm) \cite{meng2020lorentz}. 
NiTe$_2$ stands out among these materials, as the Dirac point (DP) lies in close vicinity of the Fermi level \cite{xu2018topological}. 
For all the previous materials, \textit{ab-initio} density-functional theory (DFT) calculations confirm the type-II character. 
However, the group velocity that determines the classification of the Dirac point  
is renormalized when many-body effects are included \cite{kotov2012electron,trevisanutto2008ab,banerjee2020interacting,wehling2014dirac}. 
As encountered by Kotov \textit{et al.}  \cite{kotov2012electron} and by Trevisanutto  \textit{et al.}  \cite{trevisanutto2008ab}, 
the enhancement of the electronic correlation near the weakly screened carriers at the Fermi level in graphene leads to an isotropic 
renormalization of the Fermi velocity by 17\% at the GW level,  {in better agreement with the experiment \cite{Bostwick}}. 
In type-II Dirac materials the situation is different: being highly anisotropic, we expect many-body corrections to deform the Dirac cones, 
which can lead  to a change in the topology 
if correlation effects are strong enough, as recently observed by Beau {\it et al.} in T$_d$-MoTe$_2$ with ultrafast laser pulses \cite{beaulieu2021ultrafast}. Namely, a renormalization of the electronic band structure due to dynamical changes in the effective electronic correlations caused a transient change of the spectral weight on the Fermi surface, yielding a Lifshitz transition.

The purpose of the present work is to examine the many-body effects on the band dispersion and topology of commonly studied type-II materials: NiTe$_2$, PtSe$_2$ and PtTe$_2$. 
This has been achieved by calculating the \textit{ab-initio} electron self-energy at the GW level \cite{hedin1965new}. 
We find that many-body effects have a tendency to tilt the NiTe$_2$ Dirac cone towards the type-I situation, with an enhanced and highly anisotropic Dirac fermion velocity ($v_{\rm DF}$) renormalization. Although the induced tilt is not large enough to induce a topological transition in NiTe$_2$, we observe change in the band orderings in PtSe$_2$ and PtTe$_2$, indicating a fundamental change in the topology as correlations are included.  This work underscores the importance of correlated effects on the band topology, mainly due to the effects of the $d$-bands, which is far from being obvious {\it a priori}.

\section{Type-II Dirac points and Computational methods}

\begin{figure}[!hbt]\includegraphics[width=1\linewidth]{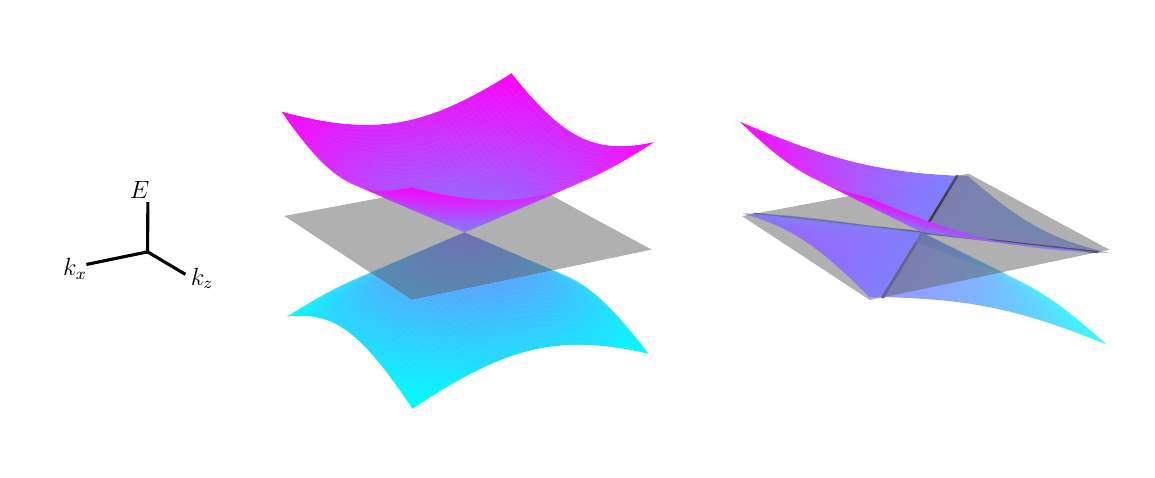}
   \caption{
In a type-I Dirac cone (left) the Fermi surface can be reduced to a point, whereas in a type II (right) 
the boundary between electron and hole pockets describes an open curve. 
}\label{fig:figPic}
\end{figure}

Whereas type-I Dirac fermions present a point-like Fermi surface, type-II Dirac fermions  emerge at the boundary  between electron and hole pockets, commonly termed as Fermi arc
[see dark lines of \cref{fig:figPic} (right)]. 
This represents a novel type of fermion that breaks Lorentz invariance due to the over-tilting of the Dirac cone [6], yielding physical properties that are very different from those of type-I Dirac electrons.
In particular, some TMDs feature type-II three-dimensional bulk Dirac fermions as a general consequence of a trigonal crystal field within the chalcogen $p$-orbitals \cite{bahramy2018ubiquitous}. 
Namely, a band-inversion on the main rotations along the growth direction  $k_z$ results in symmetry-protected crossings of the $A_1$ ($p_z$-orbital)  and $E$-derived bands ($p_x$ and $p_y$), 
which are split by a crystal field at $\Gamma$ \cite{bahramy2018ubiquitous}. These bands behave differently under the three-fold rotations $C_{3\nu}$, protecting the crossing against hybridization, which is ensured along the $\Gamma \to \mathrm{A}$ direction. 

As stated by Soluyanov {\it et al.} \cite{soluyanov2015type}, the most general Hamiltonian describing Weyl and DPs has the form:
\begin{equation}
    H(\kb)=\sum_{ij} k_i A_{ij}  \sigma_{j},
\end{equation}
where $A_{ij}$ is a generic $3 \times 4$ coefficient matrix with indices $i=x,y,z$ and $j=0,x,y,z$ and $\sigma_{j}$ corresponds to the $2 \times 2$ identity matrix ($j=0$) or the Pauli matrices ($j=x,y,z$). 
Using that $\sigma_j^2 = \sigma_0$, the spin-degenerate eigenvalue spectrum is trivially given by:
\begin{align}
    \epsilon_{\pm} (\kb) &= \sum_{i} k_i A_{i0}  \pm \sqrt{\sum_j \left( \sum_{i} k_i A_{ij} \right)^2} \nonumber
    \\  
    &= T(\kb) \pm U(\kb).
\end{align}
The first term on the right-hand side can be considered the kinetic contribution $T(\kb)$, whereas the second one {involves internal degrees of freedom and hence,} takes the role of the potential component $U(\kb)$. 
It is the former term that induces the tilting of the Dirac dispersion. 
Whenever the ratio $|R({\bm k})| = |T({\bm k})/U({\bm k})|$ is larger than one for a given $\kb$ direction, the Dirac cone is of type-II, as the tilt becomes large enough to cause contact between the 
open electron and hole pockets, in contrast to standard point-like Dirac touching (see \cref{fig:figPic}). 

In Ref. \cite{soluyanov2015type} a fit has been used to deduce the value of aforementioned ratio. We evaluate R using:
\begin{equation}
    R (\kb) =\frac{\epsilon_{+}(\kb) + \epsilon_{-}(\kb)}{\epsilon_{+}(\kb) - \epsilon_{-}(\kb)}, 
    \label{Eq:R}
\end{equation}
showing that the dispersions of the upper and lower bands are sufficient for R, eliminating the need for low-energy Hamiltonian fitting.
Note that \eqref{Eq:R} is sensitive to a rigid shift of the bands, making it necessary to set the zero-energy level to the DP. 

To obtain the band structure, we perform a ground state calculation using DFT with norm-conserving, fully-relativistic pseudopotentials in the PBE parameterization from the \texttt{PseudoDojo} library \cite{van2018pseudodojo}, using the \texttt{Quantum Espresso} package \cite{giannozzi2017advanced}.
The relaxation of the unit cell was performed until atomic forces are less than $10^{-3} \, \text{Ry/Bohr}$.
We then perform single-shot many-body calculations within the GW approximation using the \texttt{Yambo} package \cite{sangalli2019many}, which employs the  Kohn-Sham energies and wave functions from the
previous ground state calculation as input to generate the corrected quasiparticle energies and amplitudes.
Both the plasmon pole approximation and real frequency integration methods are used to evaluate the self-energy. 
Employing the quasiparticle energies and the Wannier-orbitals from the ground state calculation, an effective many-body Hamiltonian is  then generated with the \texttt{wannier90} package \cite{Pizzi2020}.
This Hamiltonian is then used to generate the quasiparticle energies in any region of the Brillouin zone (BZ) and for any band in an accurate way via the Wannier interpolation.
A cutoff of $\SI{30}{Ha}$ was needed to converge ground state properties such as the total energy and the Fermi level within the range of one meV. 
However, for the subsequent GW calculation, a cutoff of  $\SI{60}{Ha}$  and k-mesh of $14 \times 14 \times 8$ were necessary to achieve convergence  due to the presence of localized {$d$}-orbitals. 
For the screening computation, convergence has been achieved using $600$ empty bands in the RPA formulation. 
A screening cutoff as high as $\SI{11}{Ha}$ was needed to converge the quasiparticle energies and the Fermi level.

\section{Results and Discussion}


NiTe$_2$ crystallizes in the standard layered hexagonal structure of TMDs with 12 symmetry operations including inversion symmetry (space group 164). 
As a result, NiTe$_2$ possesses two symmetry-related type-II DPs  along the $\Gamma \to \mathrm{A}$ path located at $(0,0, \pm Q)$ \cite{ghosh2019observation,mukherjee2020fermi,nurmamat2021bulk}. 
Namely, the crossing bands $E$ ($p_x$, $p_y$) and $A_1$ ($p_z$)  present opposite rotation character, 
preventing hybridization and resulting in quadruply degenerate type-II DP \cite{bahramy2018ubiquitous,mukherjee2020fermi,clark2019general,yan2017lorentz,huang2016type}. 

\begin{figure}[!hbt]
\includegraphics[width=.75\linewidth]{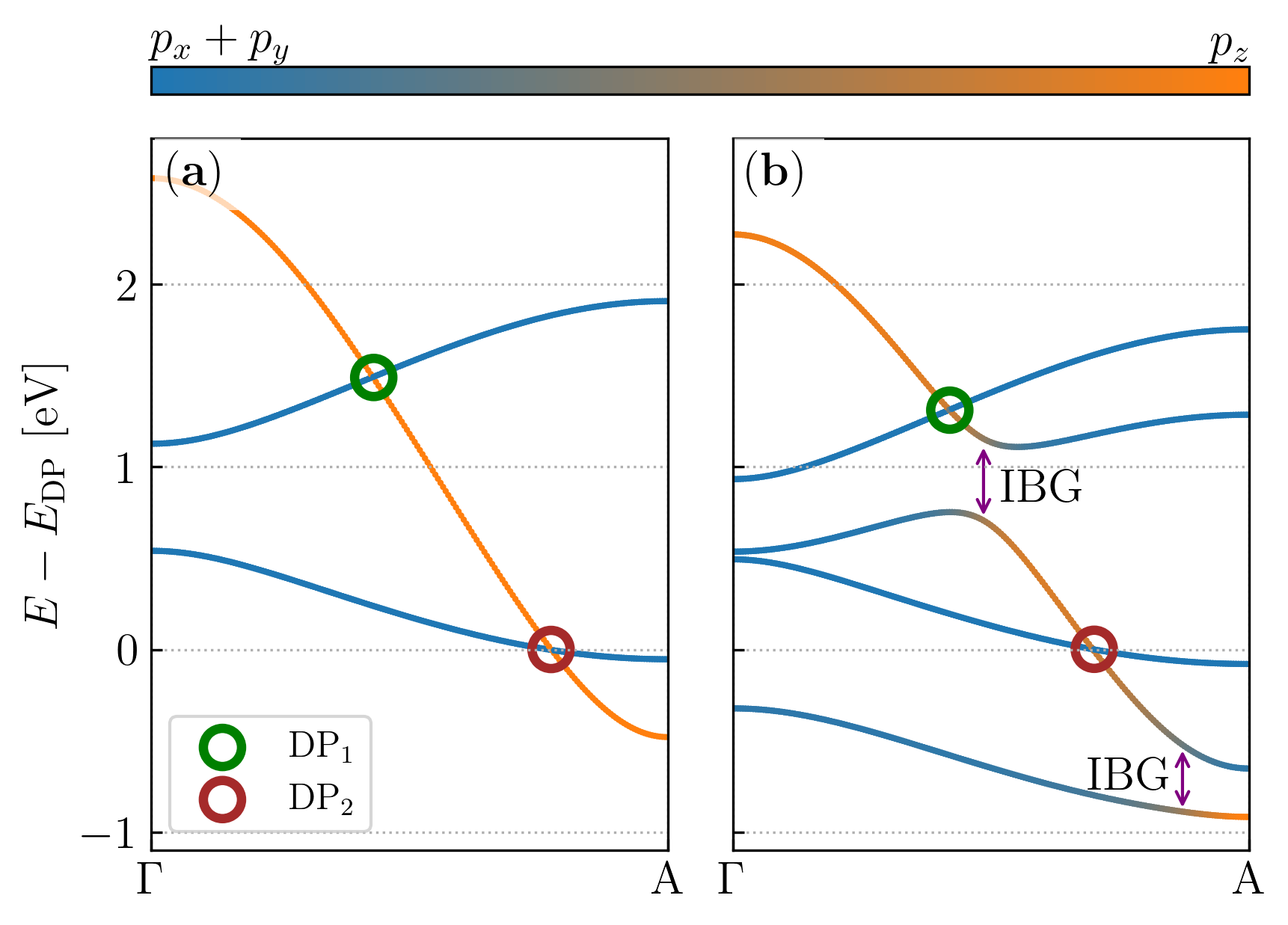}
\caption{ 
Projection of the $A_1$ and $E$ bands onto the Tellurium $p$-orbitals of NiTe$_2$, showing the $p_z$-orbital contributions (blue) and $p_{x,y}$ (orange) 
without SOC (a) and with SOC (b). DP$_1$ and DP$_2$ label the protected-by-symmetry crossings. Two IBGs are observed in (b), owing to SOC mixing. The DP$_2$ serves as the reference or zero-energy level.}
\label{fig:figProjB}
\end{figure}

In order to verify the existence of the type-II DP, we first employ a Löwdin population analysis within the \texttt{Quantum Espresso} package to project the wave 
functions onto atomic orbitals \cite{giannozzi2017advanced}.
We utilize scalar-relativistic pseudopotentials and, in a first step, neglect spin-orbit interaction to preserve $L$ and $S$ as good quantum numbers and find that the major contribution of the
crossing bands stems from the $p$-orbitals of the Te atoms. 
The splitting between the $E$ ($p_x$- and $p_y$-orbitals) and $A_1$ bands ($p_z$-orbitals) is a consequence of the crystal field splitting (CFS), which is inverted along the $\Gamma \to \mathrm{A}$ path. 
The type-II DP occurs when the CFS and the bandwidth of the dispersive $p_z$-orbitals have suitable magnitudes.
\cref{fig:figProjB}  shows the contribution of the $p_z$-orbital (blue, $A_1$ anti-bonding band) and $p_{x,y}$ ones (orange, $E$ bonding and anti-bonding)  
with (a) and without spin orbit coupling (SOC) (b), where the $E$ manifold splits further into doublets. {We observe a CFS of about 1.46 eV at $\Gamma$, whereas SOC is of the order of 360 meV (see double-headed arrows), where the latter is responsible for the splitting of the blue colored bands.}
The crossings are protected by symmetry, as the involved bands transform differently under rotations \cite{bahramy2018ubiquitous}. 
As shown in \cref{fig:figProjB}  (b) DP$_1$ results from the crossing of $|j=1/2, |m_j| = 1/2\rangle$ (original $E$ band)  and the 
upper split-off band $|j=3/2, |m_j| = 3/2\rangle$, composed mainly of the in-plane $p$-orbitals.
On the other hand, the lower SOC split-off band $|j=3/2, |m_j| = 1/2\rangle$ mixes with the $|j=1/2, |m_j| = 1/2\rangle$ band, featuring an inverted band gap (IBG), where the band character flips on either side of the avoided crossing. Note that DFT calculations using SOC and spin polarization confirms the non-magnetic ground state of NiTe$_2$ despite the presence of Ni, rendering spin polarization unnecessary.

We stress that we employ the DP energy as the reference energy, allowing a better visualization of the Dirac carrier velocity correction, which is the focus of this work. Note that existing calculations yield the  location of the DP with respect to the Fermi level at different positions: 
Ghosh {\it et al.} yield a DP $\SI{20}{meV}$ above the Fermi level \cite{ghosh2019observation}, Mukherjee {\it et al.} encounter a DP $\SI{76}{meV}$ above the Fermi level \cite{mukherjee2020fermi}, comparable to 
the value reported by  Karn {\it et al.} \cite{KARN2023}, whereas  
Nurmamat {\it et al.} report this crossing $\SI{150}{meV}$ below Fermi energy \cite{nurmamat2021bulk}. 
The discrepancy is due to the starting functionals of the DFT calculations and to the high sensitivity of the DP position to small variations of structural parameters.  Our initial DFT computation  with relaxed cell parameters
yielded a DP $\SI{50}{meV}$ above the PBE Fermi level, while it appears $\SI{0.61}{eV}$ below the Fermi energy employing G$_0$W$_0$.

\begin{figure}[!hbt]
\centering\includegraphics[width=.75\linewidth]{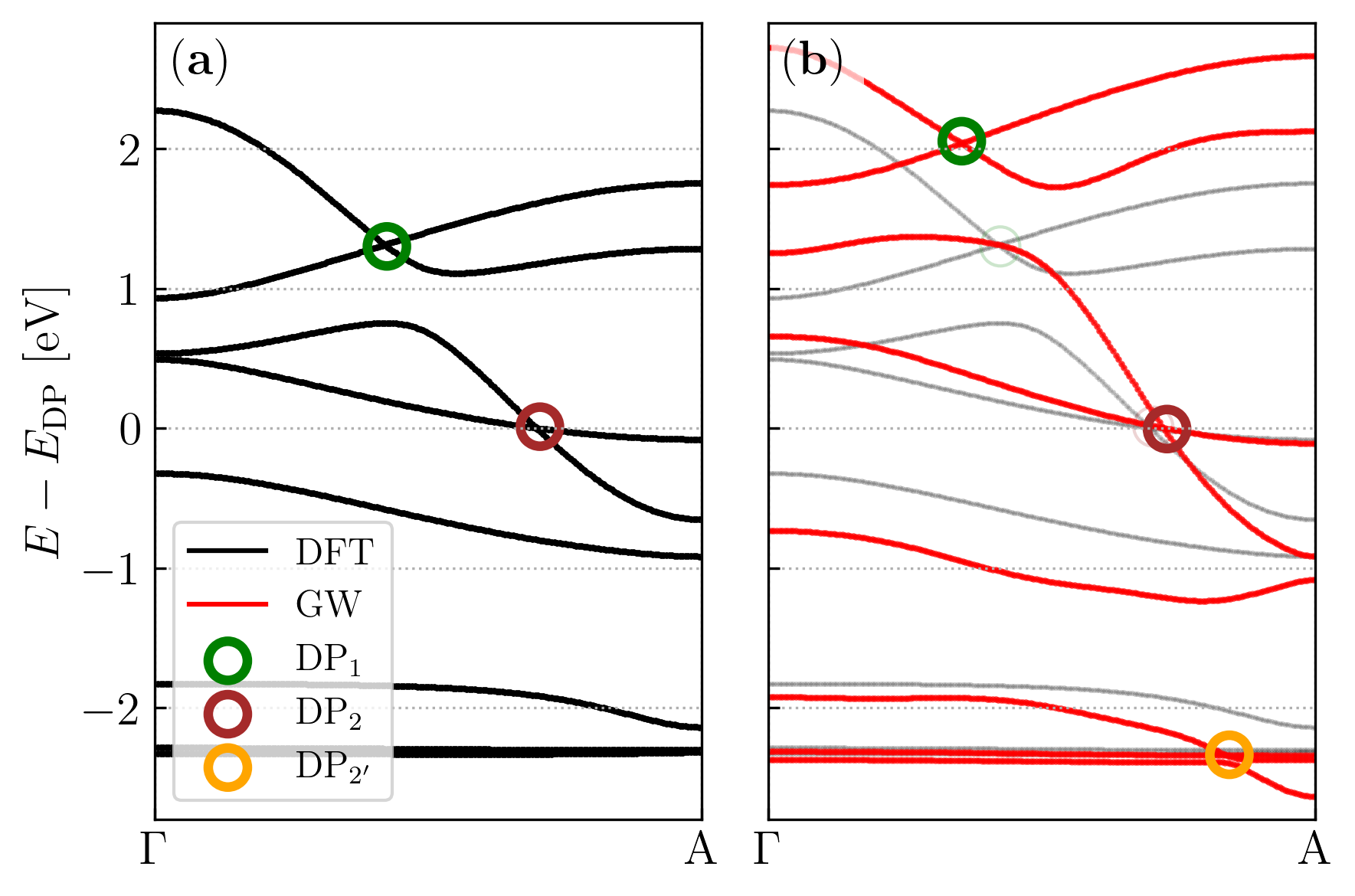}
\caption{Band structure for NiTe$_2$ along the $\Gamma \to \mathrm{A}$ direction employing DFT (a) and GW (b). The DFT results are overlaid with the GW results for direct comparison. The band structure features a type-I DP (DP$_1$), two inverted bandgaps due to SOC and a type-II DP (DP$_2$). The DP$_2$ serves as the reference or zero-energy level.
}
\label{fig:fig1}
\end{figure}

\begin{figure}
    \centering
    \includegraphics[width=.7\linewidth]{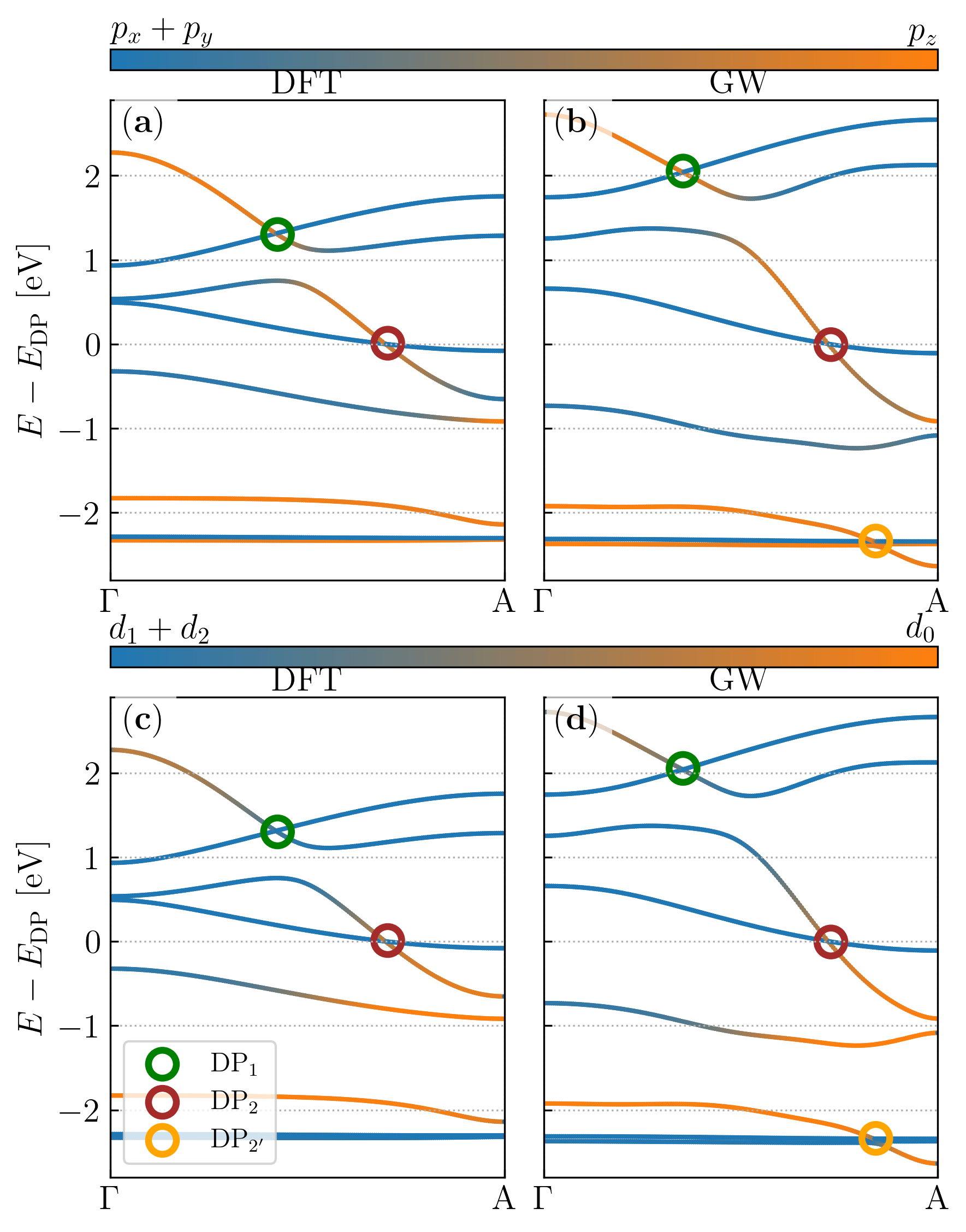}
    \caption {Normalized relative orbital projections of the $p$-orbitals in (a,b) and $d$-orbitals in (c,d) comparing DFT and GW results along the $\Gamma \to \mathrm{A}$ path for NiTe$_2$, respectively. The color indicates the relative weight of the contributions as labeled above. The circles indicate the type I (green) and type-II (brown, orange) DPs, respectively. {For more details, see  \cref{fig:op1a1})}}
    \label{fig:op1}
\end{figure}

\begin{figure}
    \centering
    \includegraphics[width=\linewidth]{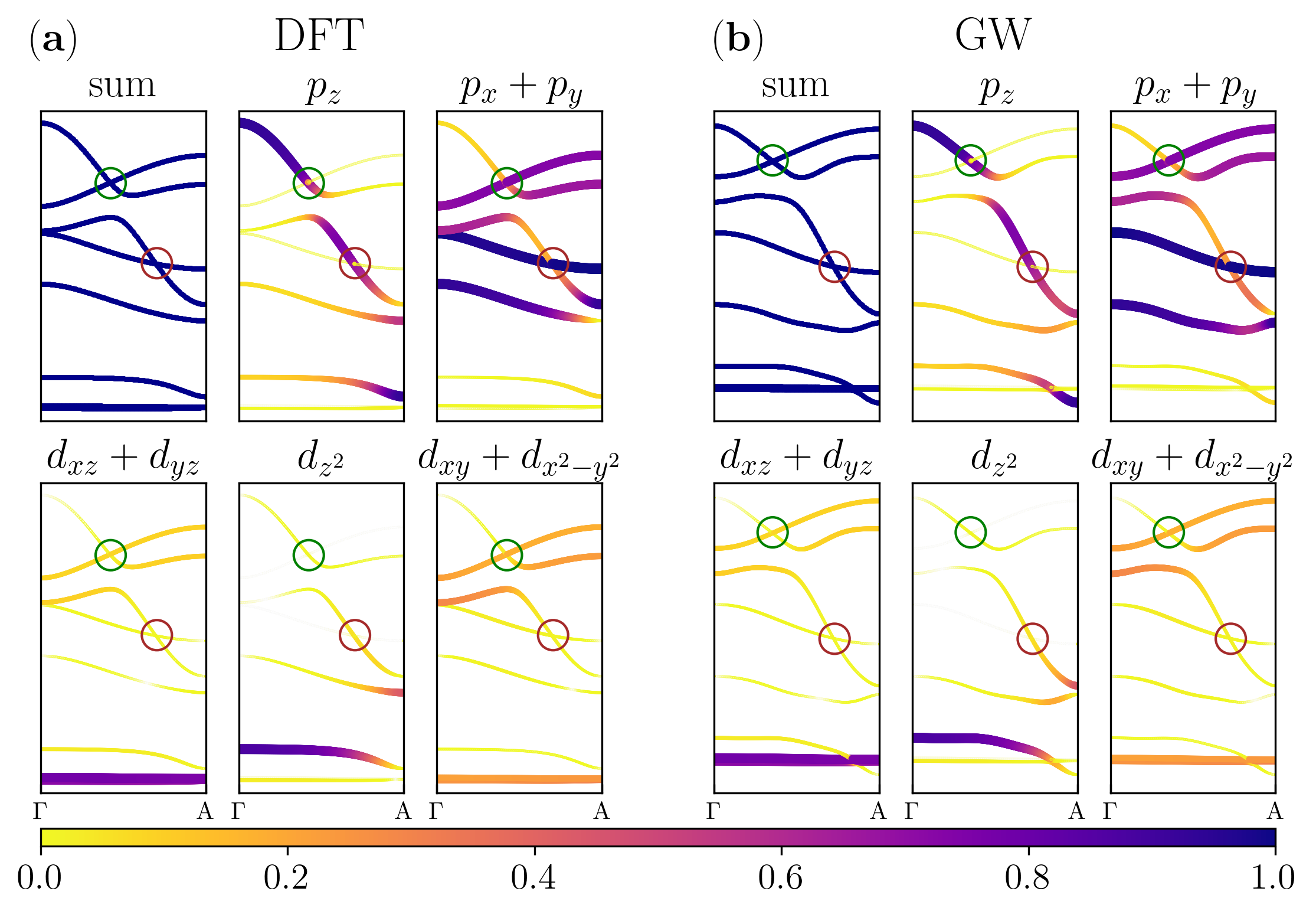}
    \caption {Normalized absolute projections of the $p$-orbitals (upper panels) and $d$-orbitals (lower panels) comparing DFT (a) and GW  (b) results along the $\Gamma \to \mathrm{A}$ path for NiTe$_2$, respectively. The color and size indicates the relative weight of the contributions as labeled above. The circles indicate the type I (green) and type-II (brown, orange) DPs, respectively. The individual panels focus on the orbitals labeled above.}
    \label{fig:op1a1}
\end{figure}

Once we have confirmed a type-II DP, we  include 
the correlations via  GW calculation at the single shot (G$_0$W$_0$). 
The CFS and band splitting are renormalized by many-body effects, and hence, the preservation of the topology cannot be guaranteed. 
\cref{fig:fig1} shows the band structures obtained with DFT (a) and G$_0$W$_0$ (b), allowing for a direct comparison.
The band structure features a type-I and a type-II DP (green and brown circles in \cref{fig:fig1}, respectively) and two IBGs. 
The GW correction enhances the effect of the localized $d$-orbitals, which can potentially modify the band ordering. To illustrate this, we include the orbital projections of the different bands in \cref{fig:op1}. 
As we can appreciate, the top five bands correspond to the $p$-orbitals of both Te atoms {(see also the top row of Fig. \cref{fig:op1a1}), 
where the absolute orbital projections are displayed}. 
Band ordering and topology are preserved on these top bands both at the DFT and GW level, showing the original type-I and type-II DP. 
However, the two lower $d$-bands that show no crossings using DFT, feature a band crossing (DP$_{2^\prime}$) when including the GW corrections, signaling that the band topology could be affected as the electronic correlations are accounted for. 

It is generally accepted that the Fermi level calculation within GW is cumbersome, as it requires corrections of the whole valence bands manifold, including  deep and very localized levels. 
The plasmon pole approximation used in this work is particularly known to be more challenging for the deep and localized states. Note that our DFT in-plane cell parameters are in very good agreement with experiments \cite{monteiro2017synthesis} with less 
than $0.01 \, \% $ discrepancy, whereas our out-of-plane crystal axis is, however, $1.25 \, \% $ shorter than the experimental one (see table \ref{tab:cell}). As Ferreira \textit{et al.} \cite{ferreira2021strain}, pointed out, the DP$_2$ energy shifts by a few hundred meV when the unit cell is strained. We indeed observed a shift of the DP$_2$ to 300 meV below the Fermi level by performing G$_0$W$_0$ calculations using the experimental c-axis length instead of the relaxed DFT value. This sets the DP just about  $150$ meV below  the experimentally obtained value \cite{nurmamat2021bulk}. Moreover, the Fermi velocity renormalization and band topology were insensitive to the cell parameters.    
\begin{table} [hbt!]
\caption{\label{tab:cell} Experimental  versus  DFT cell parameters for NiTe$_2$: lattice constant in-plane ($a$) and out-of-plane ($c$), both in \AA.}
\begin{ruledtabular}
\begin{tabular}{ccc}
         & $a$ & $c$    \\
\hline
PBE & 3.854 & 5.197   \\  
Exp. \cite{monteiro2017synthesis} & 3.858 & 5.264 
\end{tabular}
\end{ruledtabular}
\end{table} 



We examine the real frequency integration of the self-energy as implemented in the \texttt{Yambo} package. 
In particular, the zero-momentum limit of the electron energy loss spectrum, $E_{\rm LS}$ give access to
the dynamical behavior of the dielectric function $\varepsilon^{-1} (q \simeq 0,\omega)$, namely, 
\[
E_{\rm LS}( q\simeq 0, \omega) = -\Im[\varepsilon^{-1} (q \simeq 0,\omega)]. 
\]
Peaks of $E_{\rm LS} ( q\simeq 0, \omega)$ are related to plasmon resonances where $|\varepsilon(q \simeq 0,\omega) | =0$. 
We indeed observe a single pole in $\varepsilon^{-1}$ in the frequency domain, indicating the validity of the single plasmon pole approach. 
It is worth noting that the $z$-renormalization factor in our G$_0$W$_0$ calculation is close to 1 
($z \approx 0.8$) indicating a low sensitivity of the GW calculation to the starting DFT computation. 
These observations discard the possibility that the starting exchange-correlation functionals are responsible for overshooting the Fermi level.
\begin{figure}[!hbt]
\includegraphics[width=0.5\linewidth]{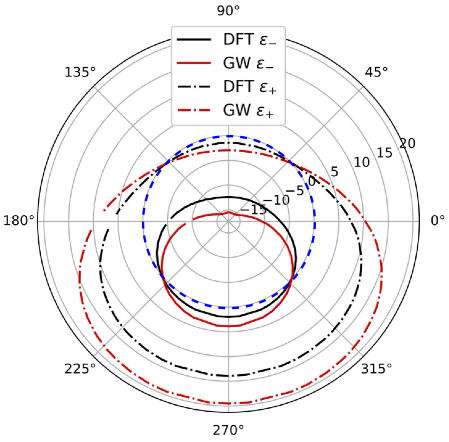} 
\caption{ 
Polar plot showing  $v_{\rm DF}$ for the $\epsilon_{\pm}$ bands forming the DP$_2$ cone, where $0^\circ$ is chosen to be the $x$-axis (blue dashed circle corresponds to zero $v_{\rm DF}$). 
The radial component is in units of 10$^6$ m/s. 
The black curves correspond to DFT results, whereas the red ones correspond to the
GW calculation. 
The dashed and solid curves correspond to the upper and lower bands, respectively.   
}
\label{fig:fig2}
\end{figure}
\begin{figure*}[htb!]
\includegraphics[width=.85\linewidth]{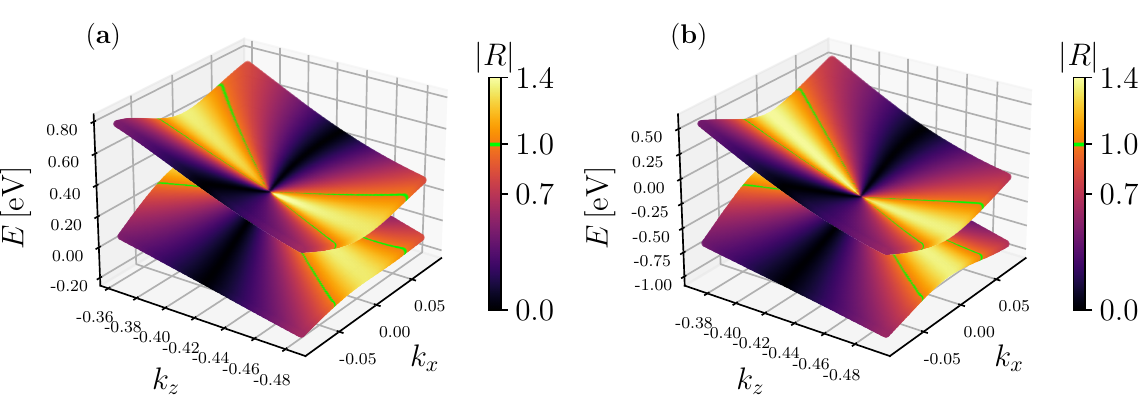}
\caption{The $\epsilon_{\pm}$ band dispersion of NiTe$_2$, forming the type-II Dirac point (DP$_2$), is visualized in 3D within the $xz$-plane for both (a) DFT and (b) GW computations. 
The color bar indicates the absolute value of the $R$ ratio (see main text). The Fermi level in the DFT results is set to zero, while the zero-energy reference for the GW energies is shifted to the DP$_2$ energy.}
\label{fig:fig3}
\end{figure*}
In order to capture the velocity renormalization anisotropy of the type-II DP, we show in \cref{fig:fig2} (a) a polar plot where the radial part indicates the Dirac fermion velocity $v_{\rm DF}$ (in units of 10$^6$ m/s) and the angle $\theta$ denotes the direction within the $xz$-plane, where we chose 0$^\circ$ and 90$^\circ$ to be the $x$- and $z$-directions, respectively. We focus on the $xz$-plane, as the  $yz$-plane can be obtained by symmetry. The black curves correspond to DFT results, whereas the red ones correspond to the 
GW calculation. Solid and dashed curves correspond to $\epsilon_-$ and $\epsilon_+$, respectively. 

We notice sign changes of $\epsilon_-$ at $-\frac{\pi}{2} \pm \frac{\pi}{4}$ and $\epsilon_+$ at $\frac{\pi}{2} \pm \frac{\pi}{4}$,
which is consistent with the type-II Dirac cone (see \cref{fig:figPic}). 
This change of sign offers thus a straightforward manner to distinguish between  type-I and type-II Dirac cones. 
We observe that, unlike along $z$, $v_{\rm DF}$ is symmetric along  the $x$-direction, which is  inherited from the unit cell structure. 
$v_{\rm DF}$ varies between $-15$ and $\SI{20e6}{\meter/\second}$, that is, the carriers are faster than in graphene for some directions in reciprocal space \cite{Elias2011}. 
The GW renormalization of $v_{\rm DF}$ is highly anisotropic and varies between $25 \, \% $ and $115\, \%$. 
This renormalization is relatively large compared to the isotropic correction of $17 \, \%$ in graphene \cite{trevisanutto2008ab,kotov2012electron}. 
This is surprising, as 
the reduced dimensionality of graphene makes the screening relatively weak (i.e., enhanced electron correlations), as compared to the 
3D bulk structure of NiTe$_2$. 
However, as stated in Ref. \cite{mukherjee2020fermi} 
the $d$-orbitals play an important role in the band structure of NiTe$_2$, yielding important GW corrections.
In our GW calculation, the influence of the $d$-orbitals is taken into account, as
the convergence of the exchange component of the self-energy requires a relatively high cutoff 
(over $120$ Ry). 
This is a clear indication of the presence of these localized $d$-orbitals with an important influence on the Dirac bands, causing 
a large GW correction. 
In particular, at the regions where $v_{\rm DF}$ is close to zero, correlation effects are enhanced due to localization in real space, 
yielding large relative corrections. 
This is evident in the region between $\theta_1 = \pi/4\  (-\pi/4) $ and $\theta_2 = 3\pi/4 \ (-3\pi/4)$ for $\epsilon_{+}$ ($\epsilon_{-}$). 
It is worth noting that GW corrections preserve the directions   where $\epsilon_\pm$ are zero, 
implying that the topology remains intact, as will be discussed below. 

\cref{fig:fig3} (a) and (b) show a 3D plot of the $\epsilon_{\pm}$ branches forming the type-II Dirac cone at both the DFT and GW level, 
where the color bar indicates the absolute value of the kinetic to potential energy ratio  $R({\bm k})$, as defined in \cref{Eq:R}. 
 This has been achieved using the Wannier interpolation techniques, as implemented in the \texttt{wannier90} code \cite{Pizzi2020} and the \texttt{WannierTools} code \cite{WU2017} on top of the DFT/GW bands.
The existence of a region with $|R({\bm k})| > 1$ is characteristic of a type-II Dirac cone \cite{soluyanov2015type}, which is observed in the BZ region near the $k_x$ axes (yellow).
Within that region, either the upper band $\epsilon_+$  (right side of DP) or the lower band $\epsilon_-$  (left side of DP) crosses the Fermi level. 
The Fermi arc, where $v_{\rm DF}$ vanishes, is highlighted with light green lines in \cref{fig:fig3}.
We note that the effect of the GW correction is to reduce the $|R({\bm k})| > 1$ region, although no Lifshitz transition occurs.

\begin{figure}[!hbt]
\includegraphics[width=.75\linewidth]{ 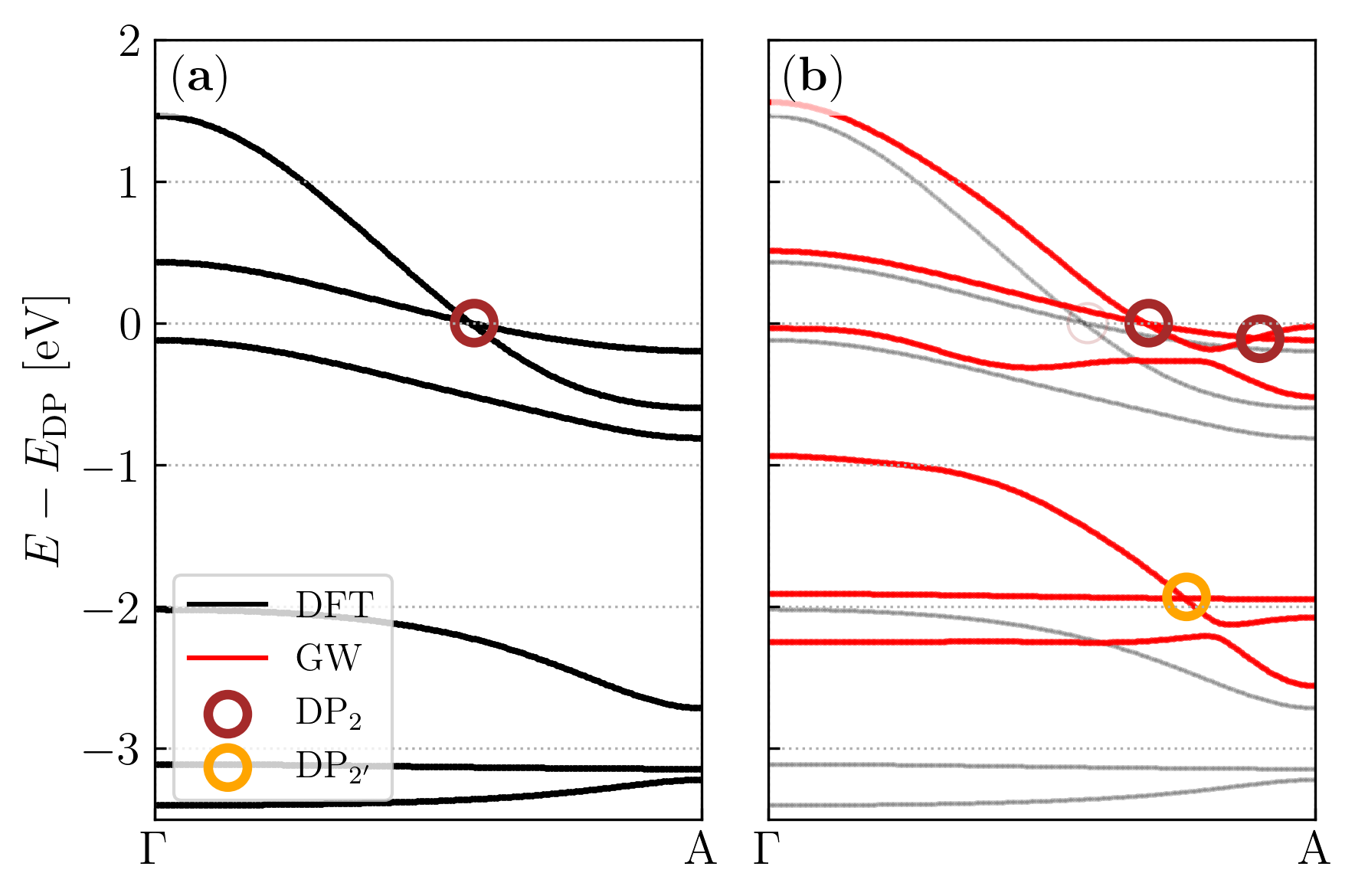} 
\caption{Band structure for PtSe$_2$ along the $\Gamma \to \mathrm{A}$ direction employing DFT (a) and GW (b). The DFT results are overlaid with the GW results for direct comparison. The type-II DPs are marked with brown and orange circles, respectively.}
\label{figPtSe2}
\end{figure}

We further investigate the effect on other type-II Dirac materials, namely PtSe$_2$ and PtTe$_2$. 
We first compute the orbital projection on PtSe$_2$ along the $\Gamma \to \mathrm{A}$ path in the absence of SOC and encounter a slightly different band ordering as compared to NiTe$_2$. 
Namely, the (anti-bonding) $E$ band 
is above the $A_1$ band, which can only cross with the bonding $E$ band as it disperses and the CFS is inverted. Note that the  $A_1$ band is less dispersive  compared to that of NiTe$_2$. This can be explained in terms of the smaller atomic radii of the chalcogen \text{($r_{\rm Se}<r_{\rm Te} $)}, which is  responsible for the out-of-plane dispersion (weaker $p_z$ bond), while the unit cell of both materials have comparable dimensions. 

We then proceed as above and compute the band structure numerically. \cref{figPtSe2} shows the DFT results and GW corrections for PtSe$_2$. As the CFS at $\Gamma$ and the bandwidth of $E$ are within the same order of magnitude, we observe a smaller velocity renormalization. The DP$_2$ crossing is observed in the DFT calculation.
Strikingly, we observe a change in the topology of the band structure as the GW corrections are included, featuring a double crossing, see \cref{figPtSe2} (b). This suggests a transition to a trivial metal topology as correlation effects are accounted for. 
\begin{figure}
    \centering
    \includegraphics[width=.75\linewidth]{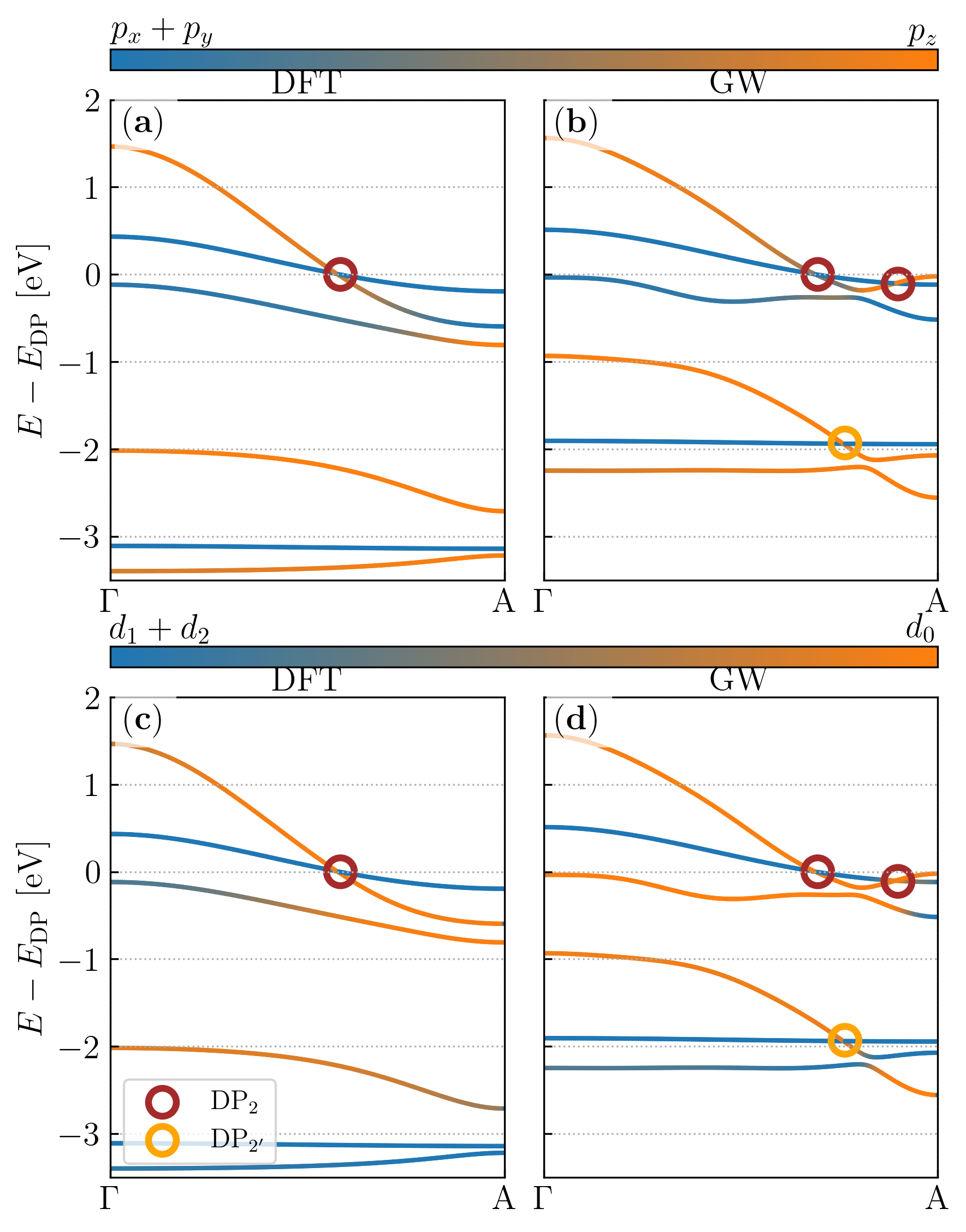}
    \caption{Normalized relative orbital projections of the $p$-orbitals in (a,b) and $d$-orbitals in (c,d) comparing DFT and GW results along the $\Gamma \to \mathrm{A}$ path for PtSe$_2$, respectively. The color indicates the relative weight of the contributions as labeled above. The circles indicate the type I (green) and type-II (brown, orange) DPs, respectively. For more details, see Fig. \ref{fig:op1a2}. }
    \label{figPtSe2OP}
\end{figure}

\begin{figure}
    \centering
    \includegraphics[width=\linewidth]{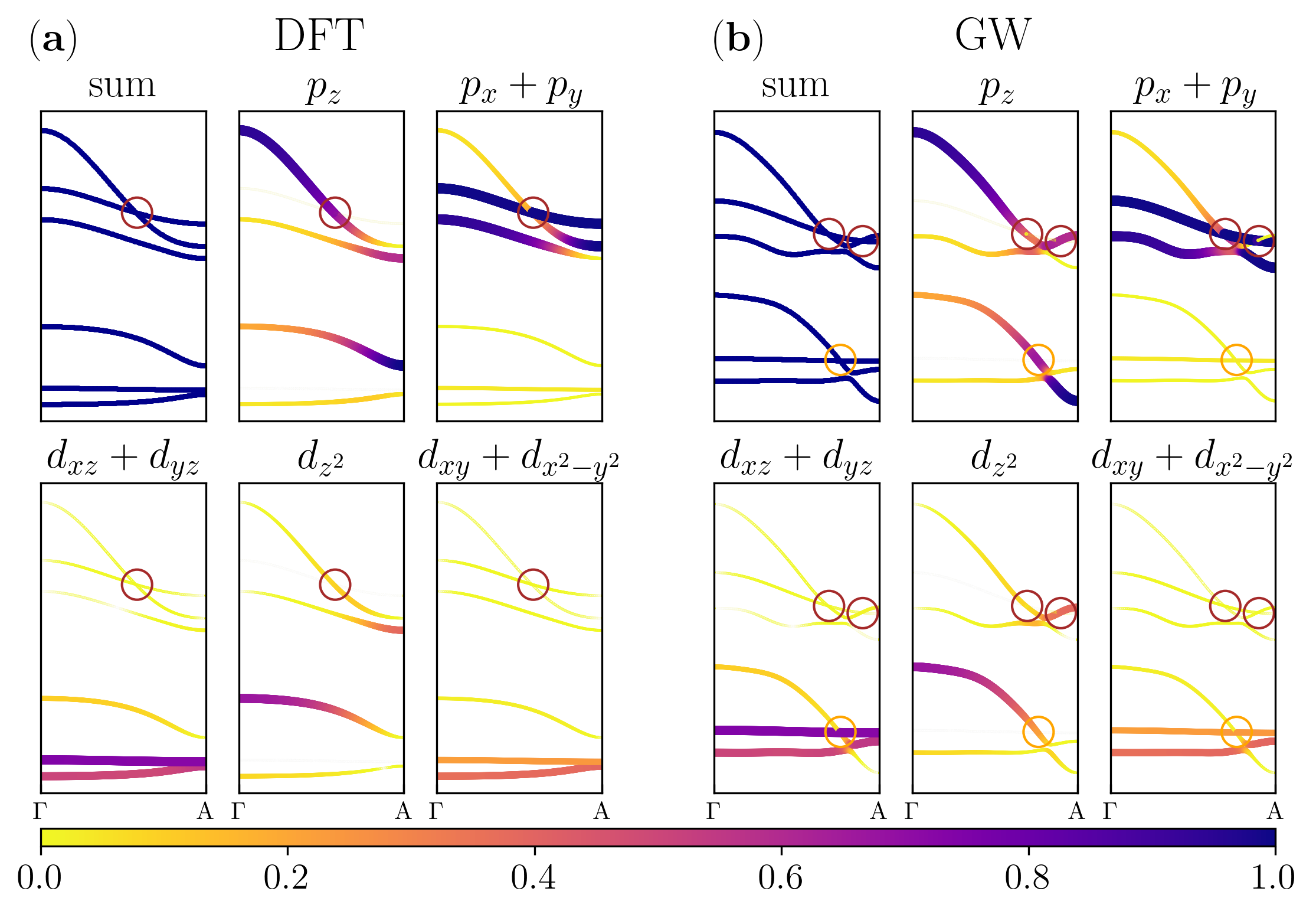}
    \caption {Normalized absoulte projections of the $p$-orbitals (upper panels) and $d$-orbitals (lower panels) comparing DFT (a) and GW  (b) results along the $\Gamma \to \mathrm{A}$ path for PtSe$_2$, respectively. The color and size  indicates the relative weight of the contributions as labeled above. The circles indicate the type I (green) and type-II (brown, orange) DPs, respectively. The individual panels focus on the orbitals labeled above.}
    \label{fig:op1a2}
\end{figure}

The orbital projections for both sets of results are illustrated in \cref{figPtSe2OP} and \cref{fig:op1a2} (normalized and relative, respectively). 
At the DFT level (left panels), we observe that the CFS inversion is maintained. However, along the $\Gamma \to \mathrm{A}$ path in the GW results, this inversion no longer occurs, indicating identical band ordering at $\Gamma$ and $\mathrm{A}$.
Conversely, we observe a crossing in the lower-lying $d$-bands at the GW level, which is absent in the DFT calculation, signaling an inversion in the CFS of the $d$ manifold.

\begin{figure}[!hbt]
\includegraphics[width=.75\linewidth]{ 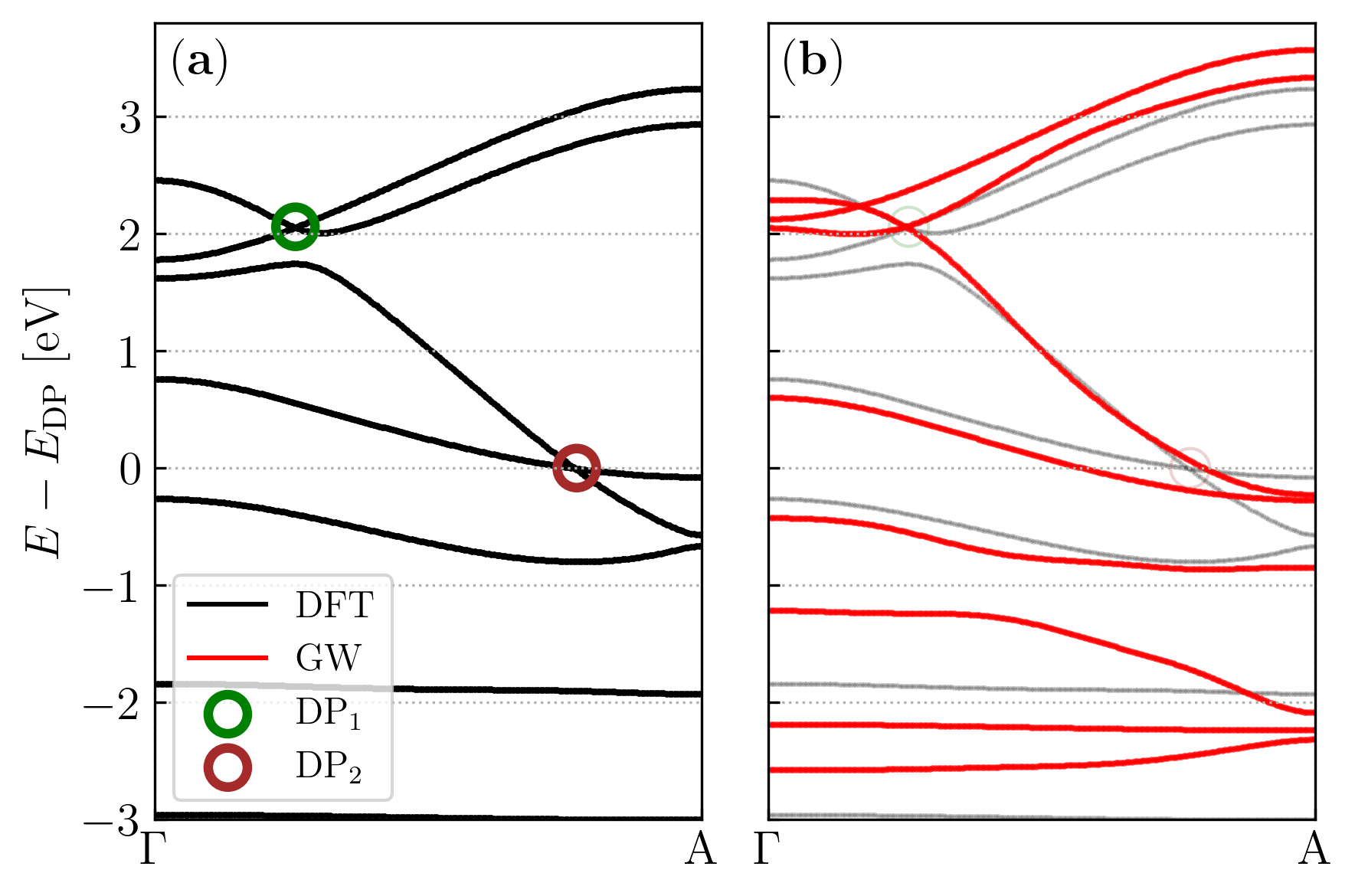}
\caption{Band structure for PtTe$_2$ along the $\Gamma \to \mathrm{A}$ direction employing DFT (a) and GW (b). The DFT results are overlaid with the GW results for direct comparison. Type-I and type-II Dirac points are denoted by green and brown circles, respectively.
}
\label{figPtTe2}
\end{figure}

We finally analyse PtTe$_2$. As \cref{figPtTe2} shows, PtTe$_2$ features a DP$_1$, IBG and DP$_2$ at the DFT level. The GW computation introduces a correction on the $E$ bands, resulting in a symmetry switching or band reordering. 
We speculate that this is due to the larger contribution of the $d$-bands, however, the exact mechanism would involve a detailed analysis of the wavefunction which is beyond the scope of this work. As a result,  we obtain two type-I DPs along the $\Gamma \to \mathrm{A}$ direction, whereas the type-II DP is absent at the GW level.  

To enhance clarity of the discussion above, we include the orbital projections with both methods, as previously done for NiTe$_2$ and PtSe$_2$. 
It is worth noting that the DP$_2$ (DFT) is a few eV below the Fermi level, rendering this material less relevant as a topologically non-trivial material, however, it underscores the nuanced influence of electronic interactions in the band structure of TMDs.  
\begin{figure}
    \centering
    \includegraphics[width=.75\linewidth]{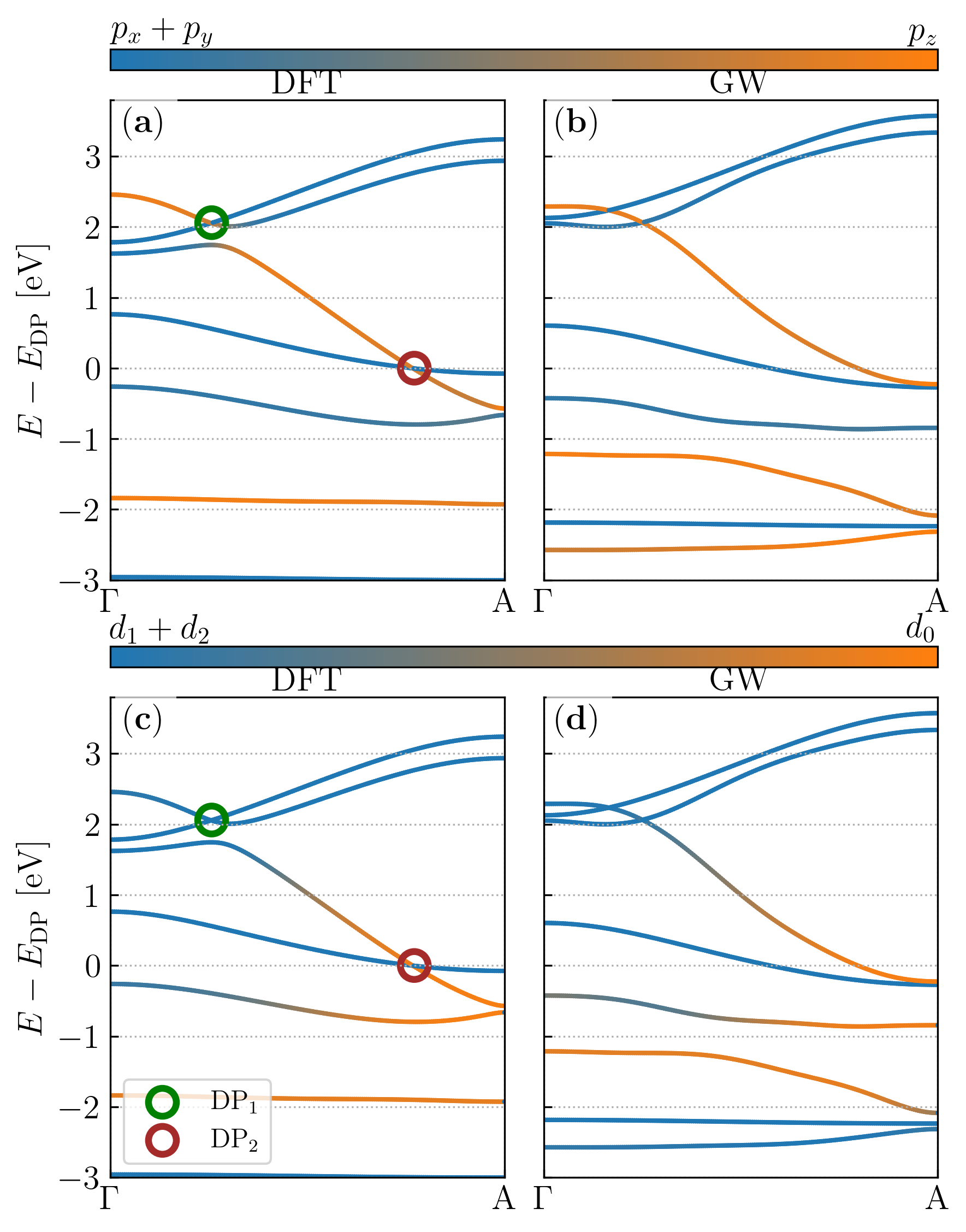}
    \caption{Normalized orbital projections of the $p$-orbitals in (a,b) and $d$-orbitals in (c,d) comparing DFT and GW results along the $\Gamma \to \mathrm{A}$ path for PtTe$_2$, respectively. The color indicates the relative weight of the contributions as labeled above. The circles indicate the type I (green) and type-II (brown, orange) DPs, respectively.}
    \label{fig:enter-label}
\end{figure}
\begin{figure}
    \centering
    \includegraphics[width=\linewidth]{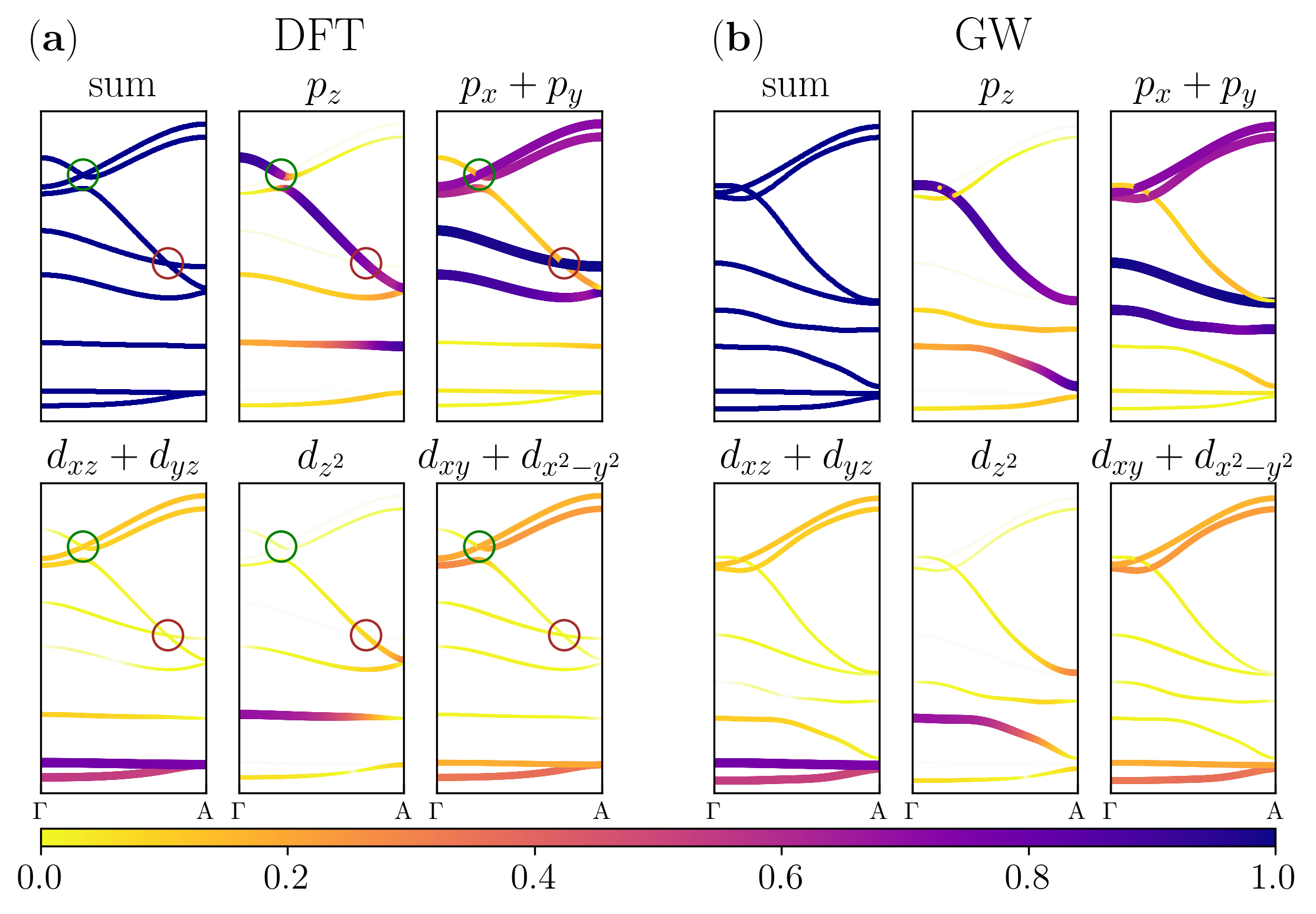}
    \caption {Normalized orbital projections of the $p$-orbitals (upper panels) and $d$-orbitals (lower panels) comparing DFT (a) and GW  (b) results along the $\Gamma \to \mathrm{A}$ path for PtTe$_2$, respectively. The color and size  indicates the relative weight of the contributions as labeled above. The circles indicate the type I (green) and type-II (brown, orange) DPs, respectively. The individual panels focus on the orbitals labeled above.}
    \label{fig:op1a3}
\end{figure}

\section{Conclusions}
In this work we have scrutinized on the effects of the GW-level many-body correlations on the band structure and topology of 
NiTe$_2$, PtSe$_2$ and PtTe$_2$. 
For the former we observe a significant renormalization of the Fermi velocity leading to a substantial tilt of the Dirac cone.
We have also examined the impact of strain and found that using the compressed DFT-relaxed unit cell can cause a shift in the energetic position of the type-II DP at both the DFT and GW levels. In contrast, using the experimentally obtained lattice constant yields better agreement of the DP energy with experimental values.
In spite of the corrections to the CFS and band bendings, the topology is preserved in  NiTe$_2$ \textit{i.e.}, it is a type-II Dirac materials both at the DFT and
at the GW level. 
{However,  PtSe$_2$ becomes a trivial metal, featuring a crossing of $d$-character bands deep below the Fermi level. Finally, we examined PtTe$_2$ and observe a double  type-I Dirac crossing, suggesting a symmetry switching as the GW corrections are introduced. These findings highlight the necessity to evaluate accurate computations, as it is not obvious \textit{a priori}}. 

\section{Acknowledgments}
This work is supported by the Cluster of Excellence 'The Hamburg Centre for Ultrafast Imaging' of the Deutsche Forschungsgemeinschaft (DFG) - EXC 1074 - project ID 194651731.
Calculations were carried out on Hummel funded by the DFG – 498394658.





\section{References}
\bibliography{paper}

\providecommand{\noopsort}[1]{}\providecommand{\singleletter}[1]{#1}%
\begin{thebibliography}{35}%
\makeatletter
\providecommand \@ifxundefined [1]{%
 \@ifx{#1\undefined}
}%
\providecommand \@ifnum [1]{%
 \ifnum #1\expandafter \@firstoftwo
 \else \expandafter \@secondoftwo
 \fi
}%
\providecommand \@ifx [1]{%
 \ifx #1\expandafter \@firstoftwo
 \else \expandafter \@secondoftwo
 \fi
}%
\providecommand \natexlab [1]{#1}%
\providecommand \enquote  [1]{``#1''}%
\providecommand \bibnamefont  [1]{#1}%
\providecommand \bibfnamefont [1]{#1}%
\providecommand \citenamefont [1]{#1}%
\providecommand \href@noop [0]{\@secondoftwo}%
\providecommand \href [0]{\begingroup \@sanitize@url \@href}%
\providecommand \@href[1]{\@@startlink{#1}\@@href}%
\providecommand \@@href[1]{\endgroup#1\@@endlink}%
\providecommand \@sanitize@url [0]{\catcode `\\12\catcode `\$12\catcode
  `\&12\catcode `\#12\catcode `\^12\catcode `\_12\catcode `\%12\relax}%
\providecommand \@@startlink[1]{}%
\providecommand \@@endlink[0]{}%
\providecommand \url  [0]{\begingroup\@sanitize@url \@url }%
\providecommand \@url [1]{\endgroup\@href {#1}{\urlprefix }}%
\providecommand \urlprefix  [0]{URL }%
\providecommand \Eprint [0]{\href }%
\providecommand \doibase [0]{https://doi.org/}%
\providecommand \selectlanguage [0]{\@gobble}%
\providecommand \bibinfo  [0]{\@secondoftwo}%
\providecommand \bibfield  [0]{\@secondoftwo}%
\providecommand \translation [1]{[#1]}%
\providecommand \BibitemOpen [0]{}%
\providecommand \bibitemStop [0]{}%
\providecommand \bibitemNoStop [0]{.\EOS\space}%
\providecommand \EOS [0]{\spacefactor3000\relax}%
\providecommand \BibitemShut  [1]{\csname bibitem#1\endcsname}%
\let\auto@bib@innerbib\@empty
\bibitem [{\citenamefont {Novoselov}\ \emph {et~al.}(2004)\citenamefont
  {Novoselov}, \citenamefont {Geim}, \citenamefont {Morozov}, \citenamefont
  {Jiang}, \citenamefont {Zhang}, \citenamefont {Dubonos}, \citenamefont
  {Grigorieva},\ and\ \citenamefont {Firsov}}]{novoselov2004electric}%
  \BibitemOpen
  \bibfield  {author} {\bibinfo {author} {\bibfnamefont {K.~S.}\ \bibnamefont
  {Novoselov}}, \bibinfo {author} {\bibfnamefont {A.~K.}\ \bibnamefont {Geim}},
  \bibinfo {author} {\bibfnamefont {S.~V.}\ \bibnamefont {Morozov}}, \bibinfo
  {author} {\bibfnamefont {D.-e.}\ \bibnamefont {Jiang}}, \bibinfo {author}
  {\bibfnamefont {Y.}~\bibnamefont {Zhang}}, \bibinfo {author} {\bibfnamefont
  {S.~V.}\ \bibnamefont {Dubonos}}, \bibinfo {author} {\bibfnamefont {I.~V.}\
  \bibnamefont {Grigorieva}},\ and\ \bibinfo {author} {\bibfnamefont {A.~A.}\
  \bibnamefont {Firsov}},\ }\bibfield  {title} {\bibinfo {title} {Electric
  field effect in atomically thin carbon films},\ }\href@noop {} {\bibfield
  {journal} {\bibinfo  {journal} {science}\ }\textbf {\bibinfo {volume}
  {306}},\ \bibinfo {pages} {666} (\bibinfo {year} {2004})}\BibitemShut
  {NoStop}%
\bibitem [{\citenamefont {Geim}\ and\ \citenamefont
  {Novoselov}(2010)}]{geim2010rise}%
  \BibitemOpen
  \bibfield  {author} {\bibinfo {author} {\bibfnamefont {A.~K.}\ \bibnamefont
  {Geim}}\ and\ \bibinfo {author} {\bibfnamefont {K.~S.}\ \bibnamefont
  {Novoselov}},\ }\bibfield  {title} {\bibinfo {title} {The rise of graphene},\
  }in\ \href@noop {} {\emph {\bibinfo {booktitle} {Nanoscience and technology:
  a collection of reviews from nature journals}}}\ (\bibinfo  {publisher}
  {World Scientific},\ \bibinfo {year} {2010})\ pp.\ \bibinfo {pages}
  {11--19}\BibitemShut {NoStop}%
\bibitem [{\citenamefont {Hasan}\ and\ \citenamefont
  {Kane}(2010)}]{hasan2010colloquium}%
  \BibitemOpen
  \bibfield  {author} {\bibinfo {author} {\bibfnamefont {M.~Z.}\ \bibnamefont
  {Hasan}}\ and\ \bibinfo {author} {\bibfnamefont {C.~L.}\ \bibnamefont
  {Kane}},\ }\bibfield  {title} {\bibinfo {title} {Colloquium: topological
  insulators},\ }\href@noop {} {\bibfield  {journal} {\bibinfo  {journal}
  {Reviews of modern physics}\ }\textbf {\bibinfo {volume} {82}},\ \bibinfo
  {pages} {3045} (\bibinfo {year} {2010})}\BibitemShut {NoStop}%
\bibitem [{\citenamefont {Giustino}\ \emph {et~al.}(2021)\citenamefont
  {Giustino}, \citenamefont {Lee}, \citenamefont {Trier}, \citenamefont
  {Bibes}, \citenamefont {Winter}, \citenamefont {Valent{\'\i}}, \citenamefont
  {Son}, \citenamefont {Taillefer}, \citenamefont {Heil}, \citenamefont
  {Figueroa} \emph {et~al.}}]{giustino20212021}%
  \BibitemOpen
  \bibfield  {author} {\bibinfo {author} {\bibfnamefont {F.}~\bibnamefont
  {Giustino}}, \bibinfo {author} {\bibfnamefont {J.~H.}\ \bibnamefont {Lee}},
  \bibinfo {author} {\bibfnamefont {F.}~\bibnamefont {Trier}}, \bibinfo
  {author} {\bibfnamefont {M.}~\bibnamefont {Bibes}}, \bibinfo {author}
  {\bibfnamefont {S.~M.}\ \bibnamefont {Winter}}, \bibinfo {author}
  {\bibfnamefont {R.}~\bibnamefont {Valent{\'\i}}}, \bibinfo {author}
  {\bibfnamefont {Y.-W.}\ \bibnamefont {Son}}, \bibinfo {author} {\bibfnamefont
  {L.}~\bibnamefont {Taillefer}}, \bibinfo {author} {\bibfnamefont
  {C.}~\bibnamefont {Heil}}, \bibinfo {author} {\bibfnamefont {A.~I.}\
  \bibnamefont {Figueroa}}, \emph {et~al.},\ }\bibfield  {title} {\bibinfo
  {title} {The 2021 quantum materials roadmap},\ }\href@noop {} {\bibfield
  {journal} {\bibinfo  {journal} {Journal of Physics: Materials}\ }\textbf
  {\bibinfo {volume} {3}},\ \bibinfo {pages} {042006} (\bibinfo {year}
  {2021})}\BibitemShut {NoStop}%
\bibitem [{\citenamefont {Wehling}\ \emph {et~al.}(2014)\citenamefont
  {Wehling}, \citenamefont {Black-Schaffer},\ and\ \citenamefont
  {Balatsky}}]{wehling2014dirac}%
  \BibitemOpen
  \bibfield  {author} {\bibinfo {author} {\bibfnamefont {T.~O.}\ \bibnamefont
  {Wehling}}, \bibinfo {author} {\bibfnamefont {A.~M.}\ \bibnamefont
  {Black-Schaffer}},\ and\ \bibinfo {author} {\bibfnamefont {A.~V.}\
  \bibnamefont {Balatsky}},\ }\bibfield  {title} {\bibinfo {title} {{Dirac}
  materials},\ }\href@noop {} {\bibfield  {journal} {\bibinfo  {journal}
  {Advances in Physics}\ }\textbf {\bibinfo {volume} {63}},\ \bibinfo {pages}
  {1} (\bibinfo {year} {2014})}\BibitemShut {NoStop}%
\bibitem [{\citenamefont {Soluyanov}\ \emph {et~al.}(2015)\citenamefont
  {Soluyanov}, \citenamefont {Gresch}, \citenamefont {Wang}, \citenamefont
  {Wu}, \citenamefont {Troyer}, \citenamefont {Dai},\ and\ \citenamefont
  {Bernevig}}]{soluyanov2015type}%
  \BibitemOpen
  \bibfield  {author} {\bibinfo {author} {\bibfnamefont {A.~A.}\ \bibnamefont
  {Soluyanov}}, \bibinfo {author} {\bibfnamefont {D.}~\bibnamefont {Gresch}},
  \bibinfo {author} {\bibfnamefont {Z.}~\bibnamefont {Wang}}, \bibinfo {author}
  {\bibfnamefont {Q.}~\bibnamefont {Wu}}, \bibinfo {author} {\bibfnamefont
  {M.}~\bibnamefont {Troyer}}, \bibinfo {author} {\bibfnamefont
  {X.}~\bibnamefont {Dai}},\ and\ \bibinfo {author} {\bibfnamefont {B.~A.}\
  \bibnamefont {Bernevig}},\ }\bibfield  {title} {\bibinfo {title} {Type-{II}
  {Weyl} semimetals},\ }\href@noop {} {\bibfield  {journal} {\bibinfo
  {journal} {Nature}\ }\textbf {\bibinfo {volume} {527}},\ \bibinfo {pages}
  {495} (\bibinfo {year} {2015})}\BibitemShut {NoStop}%
\bibitem [{\citenamefont {Udagawa}\ and\ \citenamefont
  {Bergholtz}(2016)}]{udagawa2016field}%
  \BibitemOpen
  \bibfield  {author} {\bibinfo {author} {\bibfnamefont {M.}~\bibnamefont
  {Udagawa}}\ and\ \bibinfo {author} {\bibfnamefont {E.~J.}\ \bibnamefont
  {Bergholtz}},\ }\bibfield  {title} {\bibinfo {title} {Field-selective anomaly
  and chiral mode reversal in type-{II} {Weyl} materials},\ }\href@noop {}
  {\bibfield  {journal} {\bibinfo  {journal} {Physical review letters}\
  }\textbf {\bibinfo {volume} {117}},\ \bibinfo {pages} {086401} (\bibinfo
  {year} {2016})}\BibitemShut {NoStop}%
\bibitem [{\citenamefont {Yu}\ \emph {et~al.}(2016)\citenamefont {Yu},
  \citenamefont {Yao},\ and\ \citenamefont {Yang}}]{yu2016predicted}%
  \BibitemOpen
  \bibfield  {author} {\bibinfo {author} {\bibfnamefont {Z.-M.}\ \bibnamefont
  {Yu}}, \bibinfo {author} {\bibfnamefont {Y.}~\bibnamefont {Yao}},\ and\
  \bibinfo {author} {\bibfnamefont {S.~A.}\ \bibnamefont {Yang}},\ }\bibfield
  {title} {\bibinfo {title} {Predicted unusual magnetoresponse in type-{II}
  {Weyl} semimetals},\ }\href@noop {} {\bibfield  {journal} {\bibinfo
  {journal} {Physical review letters}\ }\textbf {\bibinfo {volume} {117}},\
  \bibinfo {pages} {077202} (\bibinfo {year} {2016})}\BibitemShut {NoStop}%
\bibitem [{\citenamefont {O’Brien}\ \emph {et~al.}(2016)\citenamefont
  {O’Brien}, \citenamefont {Diez},\ and\ \citenamefont
  {Beenakker}}]{o2016magnetic}%
  \BibitemOpen
  \bibfield  {author} {\bibinfo {author} {\bibfnamefont {T.}~\bibnamefont
  {O’Brien}}, \bibinfo {author} {\bibfnamefont {M.}~\bibnamefont {Diez}},\
  and\ \bibinfo {author} {\bibfnamefont {C.}~\bibnamefont {Beenakker}},\
  }\bibfield  {title} {\bibinfo {title} {Magnetic breakdown and klein tunneling
  in a type-{II} {Weyl} semimetal},\ }\href@noop {} {\bibfield  {journal}
  {\bibinfo  {journal} {Physical review letters}\ }\textbf {\bibinfo {volume}
  {116}},\ \bibinfo {pages} {236401} (\bibinfo {year} {2016})}\BibitemShut
  {NoStop}%
\bibitem [{\citenamefont {Volovik}(2016)}]{volovik2016black}%
  \BibitemOpen
  \bibfield  {author} {\bibinfo {author} {\bibfnamefont {G.~E.}\ \bibnamefont
  {Volovik}},\ }\bibfield  {title} {\bibinfo {title} {Black hole and hawking
  radiation by type-{II} {Weyl} fermions},\ }\href@noop {} {\bibfield
  {journal} {\bibinfo  {journal} {JETP letters}\ }\textbf {\bibinfo {volume}
  {104}},\ \bibinfo {pages} {645} (\bibinfo {year} {2016})}\BibitemShut
  {NoStop}%
\bibitem [{\citenamefont {Huang}\ \emph {et~al.}(2016)\citenamefont {Huang},
  \citenamefont {Zhou},\ and\ \citenamefont {Duan}}]{huang2016type}%
  \BibitemOpen
  \bibfield  {author} {\bibinfo {author} {\bibfnamefont {H.}~\bibnamefont
  {Huang}}, \bibinfo {author} {\bibfnamefont {S.}~\bibnamefont {Zhou}},\ and\
  \bibinfo {author} {\bibfnamefont {W.}~\bibnamefont {Duan}},\ }\bibfield
  {title} {\bibinfo {title} {Type-{II} {Dirac} fermions in the {PtSe$_2$} class
  of transition metal dichalcogenides},\ }\href@noop {} {\bibfield  {journal}
  {\bibinfo  {journal} {Physical Review B}\ }\textbf {\bibinfo {volume} {94}},\
  \bibinfo {pages} {121117} (\bibinfo {year} {2016})}\BibitemShut {NoStop}%
\bibitem [{\citenamefont {Yan}\ \emph {et~al.}(2017)\citenamefont {Yan},
  \citenamefont {Huang}, \citenamefont {Zhang}, \citenamefont {Wang},
  \citenamefont {Yao}, \citenamefont {Deng}, \citenamefont {Wan}, \citenamefont
  {Zhang}, \citenamefont {Arita}, \citenamefont {Yang} \emph
  {et~al.}}]{yan2017lorentz}%
  \BibitemOpen
  \bibfield  {author} {\bibinfo {author} {\bibfnamefont {M.}~\bibnamefont
  {Yan}}, \bibinfo {author} {\bibfnamefont {H.}~\bibnamefont {Huang}}, \bibinfo
  {author} {\bibfnamefont {K.}~\bibnamefont {Zhang}}, \bibinfo {author}
  {\bibfnamefont {E.}~\bibnamefont {Wang}}, \bibinfo {author} {\bibfnamefont
  {W.}~\bibnamefont {Yao}}, \bibinfo {author} {\bibfnamefont {K.}~\bibnamefont
  {Deng}}, \bibinfo {author} {\bibfnamefont {G.}~\bibnamefont {Wan}}, \bibinfo
  {author} {\bibfnamefont {H.}~\bibnamefont {Zhang}}, \bibinfo {author}
  {\bibfnamefont {M.}~\bibnamefont {Arita}}, \bibinfo {author} {\bibfnamefont
  {H.}~\bibnamefont {Yang}}, \emph {et~al.},\ }\bibfield  {title} {\bibinfo
  {title} {Lorentz-violating type-{II} {Dirac} fermions in transition metal
  dichalcogenide {PtTe$_2$}},\ }\href@noop {} {\bibfield  {journal} {\bibinfo
  {journal} {Nature communications}\ }\textbf {\bibinfo {volume} {8}},\
  \bibinfo {pages} {1} (\bibinfo {year} {2017})}\BibitemShut {NoStop}%
\bibitem [{\citenamefont {Chang}\ \emph {et~al.}(2017)\citenamefont {Chang},
  \citenamefont {Xu}, \citenamefont {Sanchez}, \citenamefont {Tsai},
  \citenamefont {Huang}, \citenamefont {Chang}, \citenamefont {Hsu},
  \citenamefont {Bian}, \citenamefont {Belopolski}, \citenamefont {Yu} \emph
  {et~al.}}]{chang2017type}%
  \BibitemOpen
  \bibfield  {author} {\bibinfo {author} {\bibfnamefont {T.-R.}\ \bibnamefont
  {Chang}}, \bibinfo {author} {\bibfnamefont {S.-Y.}\ \bibnamefont {Xu}},
  \bibinfo {author} {\bibfnamefont {D.~S.}\ \bibnamefont {Sanchez}}, \bibinfo
  {author} {\bibfnamefont {W.-F.}\ \bibnamefont {Tsai}}, \bibinfo {author}
  {\bibfnamefont {S.-M.}\ \bibnamefont {Huang}}, \bibinfo {author}
  {\bibfnamefont {G.}~\bibnamefont {Chang}}, \bibinfo {author} {\bibfnamefont
  {C.-H.}\ \bibnamefont {Hsu}}, \bibinfo {author} {\bibfnamefont
  {G.}~\bibnamefont {Bian}}, \bibinfo {author} {\bibfnamefont {I.}~\bibnamefont
  {Belopolski}}, \bibinfo {author} {\bibfnamefont {Z.-M.}\ \bibnamefont {Yu}},
  \emph {et~al.},\ }\bibfield  {title} {\bibinfo {title} {Type-ii
  symmetry-protected topological dirac semimetals},\ }\href@noop {} {\bibfield
  {journal} {\bibinfo  {journal} {Physical review letters}\ }\textbf {\bibinfo
  {volume} {119}},\ \bibinfo {pages} {026404} (\bibinfo {year}
  {2017})}\BibitemShut {NoStop}%
\bibitem [{\citenamefont {Meng}\ \emph {et~al.}(2020)\citenamefont {Meng},
  \citenamefont {Zhang}, \citenamefont {Liu}, \citenamefont {Dai},\ and\
  \citenamefont {Liu}}]{meng2020lorentz}%
  \BibitemOpen
  \bibfield  {author} {\bibinfo {author} {\bibfnamefont {W.}~\bibnamefont
  {Meng}}, \bibinfo {author} {\bibfnamefont {X.}~\bibnamefont {Zhang}},
  \bibinfo {author} {\bibfnamefont {Y.}~\bibnamefont {Liu}}, \bibinfo {author}
  {\bibfnamefont {X.}~\bibnamefont {Dai}},\ and\ \bibinfo {author}
  {\bibfnamefont {G.}~\bibnamefont {Liu}},\ }\bibfield  {title} {\bibinfo
  {title} {Lorentz-violating type-ii dirac fermions in full-heusler compounds
  xmg2ag (x= pr, nd, sm)},\ }\href@noop {} {\bibfield  {journal} {\bibinfo
  {journal} {New Journal of Physics}\ }\textbf {\bibinfo {volume} {22}},\
  \bibinfo {pages} {073061} (\bibinfo {year} {2020})}\BibitemShut {NoStop}%
\bibitem [{\citenamefont {Xu}\ \emph {et~al.}(2018)\citenamefont {Xu},
  \citenamefont {Li}, \citenamefont {Jiao}, \citenamefont {Zhou}, \citenamefont
  {Qian}, \citenamefont {Sankar}, \citenamefont {Zhigadlo}, \citenamefont {Qi},
  \citenamefont {Qian}, \citenamefont {Chou} \emph
  {et~al.}}]{xu2018topological}%
  \BibitemOpen
  \bibfield  {author} {\bibinfo {author} {\bibfnamefont {C.}~\bibnamefont
  {Xu}}, \bibinfo {author} {\bibfnamefont {B.}~\bibnamefont {Li}}, \bibinfo
  {author} {\bibfnamefont {W.}~\bibnamefont {Jiao}}, \bibinfo {author}
  {\bibfnamefont {W.}~\bibnamefont {Zhou}}, \bibinfo {author} {\bibfnamefont
  {B.}~\bibnamefont {Qian}}, \bibinfo {author} {\bibfnamefont {R.}~\bibnamefont
  {Sankar}}, \bibinfo {author} {\bibfnamefont {N.~D.}\ \bibnamefont
  {Zhigadlo}}, \bibinfo {author} {\bibfnamefont {Y.}~\bibnamefont {Qi}},
  \bibinfo {author} {\bibfnamefont {D.}~\bibnamefont {Qian}}, \bibinfo {author}
  {\bibfnamefont {F.-C.}\ \bibnamefont {Chou}}, \emph {et~al.},\ }\bibfield
  {title} {\bibinfo {title} {Topological type-{II} {Dirac} fermions approaching
  the fermi level in a transition metal dichalcogenide {NiTe$_2$}},\
  }\href@noop {} {\bibfield  {journal} {\bibinfo  {journal} {Chemistry of
  materials}\ }\textbf {\bibinfo {volume} {30}},\ \bibinfo {pages} {4823}
  (\bibinfo {year} {2018})}\BibitemShut {NoStop}%
\bibitem [{\citenamefont {Kotov}\ \emph {et~al.}(2012)\citenamefont {Kotov},
  \citenamefont {Uchoa}, \citenamefont {Pereira}, \citenamefont {Guinea},\ and\
  \citenamefont {Neto}}]{kotov2012electron}%
  \BibitemOpen
  \bibfield  {author} {\bibinfo {author} {\bibfnamefont {V.~N.}\ \bibnamefont
  {Kotov}}, \bibinfo {author} {\bibfnamefont {B.}~\bibnamefont {Uchoa}},
  \bibinfo {author} {\bibfnamefont {V.~M.}\ \bibnamefont {Pereira}}, \bibinfo
  {author} {\bibfnamefont {F.}~\bibnamefont {Guinea}},\ and\ \bibinfo {author}
  {\bibfnamefont {A.~C.}\ \bibnamefont {Neto}},\ }\bibfield  {title} {\bibinfo
  {title} {Electron-electron interactions in graphene: Current status and
  perspectives},\ }\href@noop {} {\bibfield  {journal} {\bibinfo  {journal}
  {Reviews of Modern Physics}\ }\textbf {\bibinfo {volume} {84}},\ \bibinfo
  {pages} {1067} (\bibinfo {year} {2012})}\BibitemShut {NoStop}%
\bibitem [{\citenamefont {Trevisanutto}\ \emph {et~al.}(2008)\citenamefont
  {Trevisanutto}, \citenamefont {Giorgetti}, \citenamefont {Reining},
  \citenamefont {Ladisa},\ and\ \citenamefont {Olevano}}]{trevisanutto2008ab}%
  \BibitemOpen
  \bibfield  {author} {\bibinfo {author} {\bibfnamefont {P.~E.}\ \bibnamefont
  {Trevisanutto}}, \bibinfo {author} {\bibfnamefont {C.}~\bibnamefont
  {Giorgetti}}, \bibinfo {author} {\bibfnamefont {L.}~\bibnamefont {Reining}},
  \bibinfo {author} {\bibfnamefont {M.}~\bibnamefont {Ladisa}},\ and\ \bibinfo
  {author} {\bibfnamefont {V.}~\bibnamefont {Olevano}},\ }\bibfield  {title}
  {\bibinfo {title} {{\textit{Ab-Initio}} {G-W} many-body effects in
  graphene},\ }\href@noop {} {\bibfield  {journal} {\bibinfo  {journal}
  {Physical review letters}\ }\textbf {\bibinfo {volume} {101}},\ \bibinfo
  {pages} {226405} (\bibinfo {year} {2008})}\BibitemShut {NoStop}%
\bibitem [{\citenamefont {Banerjee}\ \emph {et~al.}(2020)\citenamefont
  {Banerjee}, \citenamefont {Abergel}, \citenamefont {{\AA}gren}, \citenamefont
  {Aeppli},\ and\ \citenamefont {Balatsky}}]{banerjee2020interacting}%
  \BibitemOpen
  \bibfield  {author} {\bibinfo {author} {\bibfnamefont {S.}~\bibnamefont
  {Banerjee}}, \bibinfo {author} {\bibfnamefont {D.~S.}\ \bibnamefont
  {Abergel}}, \bibinfo {author} {\bibfnamefont {H.}~\bibnamefont {{\AA}gren}},
  \bibinfo {author} {\bibfnamefont {G.}~\bibnamefont {Aeppli}},\ and\ \bibinfo
  {author} {\bibfnamefont {A.~V.}\ \bibnamefont {Balatsky}},\ }\bibfield
  {title} {\bibinfo {title} {Interacting {Dirac} materials},\ }\href@noop {}
  {\bibfield  {journal} {\bibinfo  {journal} {Journal of Physics: Condensed
  Matter}\ }\textbf {\bibinfo {volume} {32}},\ \bibinfo {pages} {405603}
  (\bibinfo {year} {2020})}\BibitemShut {NoStop}%
\bibitem [{\citenamefont {Bostwick}\ \emph {et~al.}(2007)\citenamefont
  {Bostwick}, \citenamefont {Ohta}, \citenamefont {Seyller},\ and\
  \citenamefont {Horn K~Rotenberg}}]{Bostwick}%
  \BibitemOpen
  \bibfield  {author} {\bibinfo {author} {\bibfnamefont {A.}~\bibnamefont
  {Bostwick}}, \bibinfo {author} {\bibfnamefont {T.}~\bibnamefont {Ohta}},
  \bibinfo {author} {\bibfnamefont {T.}~\bibnamefont {Seyller}},\ and\ \bibinfo
  {author} {\bibfnamefont {E.}~\bibnamefont {Horn K~Rotenberg}},\ }\bibfield
  {title} {\bibinfo {title} {Quasiparticle dynamics in graphene},\ }\href@noop
  {} {\bibfield  {journal} {\bibinfo  {journal} {Nature Physics}\ }\textbf
  {\bibinfo {volume} {3}},\ \bibinfo {pages} {36} (\bibinfo {year}
  {2007})}\BibitemShut {NoStop}%
\bibitem [{\citenamefont {Beaulieu}\ \emph {et~al.}(2021)\citenamefont
  {Beaulieu}, \citenamefont {Dong}, \citenamefont {Tancogne-Dejean},
  \citenamefont {Dendzik}, \citenamefont {Pincelli}, \citenamefont {Maklar},
  \citenamefont {Xian}, \citenamefont {Sentef}, \citenamefont {Wolf},
  \citenamefont {Rubio} \emph {et~al.}}]{beaulieu2021ultrafast}%
  \BibitemOpen
  \bibfield  {author} {\bibinfo {author} {\bibfnamefont {S.}~\bibnamefont
  {Beaulieu}}, \bibinfo {author} {\bibfnamefont {S.}~\bibnamefont {Dong}},
  \bibinfo {author} {\bibfnamefont {N.}~\bibnamefont {Tancogne-Dejean}},
  \bibinfo {author} {\bibfnamefont {M.}~\bibnamefont {Dendzik}}, \bibinfo
  {author} {\bibfnamefont {T.}~\bibnamefont {Pincelli}}, \bibinfo {author}
  {\bibfnamefont {J.}~\bibnamefont {Maklar}}, \bibinfo {author} {\bibfnamefont
  {R.~P.}\ \bibnamefont {Xian}}, \bibinfo {author} {\bibfnamefont {M.~A.}\
  \bibnamefont {Sentef}}, \bibinfo {author} {\bibfnamefont {M.}~\bibnamefont
  {Wolf}}, \bibinfo {author} {\bibfnamefont {A.}~\bibnamefont {Rubio}}, \emph
  {et~al.},\ }\bibfield  {title} {\bibinfo {title} {Ultrafast dynamical
  {Lifshitz} transition},\ }\href@noop {} {\bibfield  {journal} {\bibinfo
  {journal} {Science Advances}\ }\textbf {\bibinfo {volume} {7}},\ \bibinfo
  {pages} {eabd9275} (\bibinfo {year} {2021})}\BibitemShut {NoStop}%
\bibitem [{\citenamefont {Hedin}(1965)}]{hedin1965new}%
  \BibitemOpen
  \bibfield  {author} {\bibinfo {author} {\bibfnamefont {L.}~\bibnamefont
  {Hedin}},\ }\bibfield  {title} {\bibinfo {title} {New method for calculating
  the one-particle {Green}'s function with application to the electron-gas
  problem},\ }\href@noop {} {\bibfield  {journal} {\bibinfo  {journal}
  {Physical Review}\ }\textbf {\bibinfo {volume} {139}},\ \bibinfo {pages}
  {A796} (\bibinfo {year} {1965})}\BibitemShut {NoStop}%
\bibitem [{\citenamefont {Bahramy}\ \emph {et~al.}(2018)\citenamefont
  {Bahramy}, \citenamefont {Clark}, \citenamefont {Yang}, \citenamefont {Feng},
  \citenamefont {Bawden}, \citenamefont {Riley}, \citenamefont {Markovi{\'c}},
  \citenamefont {Mazzola}, \citenamefont {Sunko}, \citenamefont {Biswas} \emph
  {et~al.}}]{bahramy2018ubiquitous}%
  \BibitemOpen
  \bibfield  {author} {\bibinfo {author} {\bibfnamefont {M.}~\bibnamefont
  {Bahramy}}, \bibinfo {author} {\bibfnamefont {O.}~\bibnamefont {Clark}},
  \bibinfo {author} {\bibfnamefont {B.-J.}\ \bibnamefont {Yang}}, \bibinfo
  {author} {\bibfnamefont {J.}~\bibnamefont {Feng}}, \bibinfo {author}
  {\bibfnamefont {L.}~\bibnamefont {Bawden}}, \bibinfo {author} {\bibfnamefont
  {J.}~\bibnamefont {Riley}}, \bibinfo {author} {\bibfnamefont
  {I.}~\bibnamefont {Markovi{\'c}}}, \bibinfo {author} {\bibfnamefont
  {F.}~\bibnamefont {Mazzola}}, \bibinfo {author} {\bibfnamefont
  {V.}~\bibnamefont {Sunko}}, \bibinfo {author} {\bibfnamefont
  {D.}~\bibnamefont {Biswas}}, \emph {et~al.},\ }\bibfield  {title} {\bibinfo
  {title} {Ubiquitous formation of bulk {Dirac} cones and topological surface
  states from a single orbital manifold in transition-metal dichalcogenides},\
  }\href@noop {} {\bibfield  {journal} {\bibinfo  {journal} {Nature materials}\
  }\textbf {\bibinfo {volume} {17}},\ \bibinfo {pages} {21} (\bibinfo {year}
  {2018})}\BibitemShut {NoStop}%
\bibitem [{\citenamefont {Van~Setten}\ \emph {et~al.}(2018)\citenamefont
  {Van~Setten}, \citenamefont {Giantomassi}, \citenamefont {Bousquet},
  \citenamefont {Verstraete}, \citenamefont {Hamann}, \citenamefont {Gonze},\
  and\ \citenamefont {Rignanese}}]{van2018pseudodojo}%
  \BibitemOpen
  \bibfield  {author} {\bibinfo {author} {\bibfnamefont {M.}~\bibnamefont
  {Van~Setten}}, \bibinfo {author} {\bibfnamefont {M.}~\bibnamefont
  {Giantomassi}}, \bibinfo {author} {\bibfnamefont {E.}~\bibnamefont
  {Bousquet}}, \bibinfo {author} {\bibfnamefont {M.~J.}\ \bibnamefont
  {Verstraete}}, \bibinfo {author} {\bibfnamefont {D.~R.}\ \bibnamefont
  {Hamann}}, \bibinfo {author} {\bibfnamefont {X.}~\bibnamefont {Gonze}},\ and\
  \bibinfo {author} {\bibfnamefont {G.-M.}\ \bibnamefont {Rignanese}},\
  }\bibfield  {title} {\bibinfo {title} {The pseudodojo: Training and grading a
  85 element optimized norm-conserving pseudopotential table},\ }\href@noop {}
  {\bibfield  {journal} {\bibinfo  {journal} {Computer Physics Communications}\
  }\textbf {\bibinfo {volume} {226}},\ \bibinfo {pages} {39} (\bibinfo {year}
  {2018})}\BibitemShut {NoStop}%
\bibitem [{\citenamefont {Giannozzi}\ \emph {et~al.}(2017)\citenamefont
  {Giannozzi}, \citenamefont {Andreussi}, \citenamefont {Brumme}, \citenamefont
  {Bunau}, \citenamefont {Nardelli}, \citenamefont {Calandra}, \citenamefont
  {Car}, \citenamefont {Cavazzoni}, \citenamefont {Ceresoli}, \citenamefont
  {Cococcioni} \emph {et~al.}}]{giannozzi2017advanced}%
  \BibitemOpen
  \bibfield  {author} {\bibinfo {author} {\bibfnamefont {P.}~\bibnamefont
  {Giannozzi}}, \bibinfo {author} {\bibfnamefont {O.}~\bibnamefont
  {Andreussi}}, \bibinfo {author} {\bibfnamefont {T.}~\bibnamefont {Brumme}},
  \bibinfo {author} {\bibfnamefont {O.}~\bibnamefont {Bunau}}, \bibinfo
  {author} {\bibfnamefont {M.~B.}\ \bibnamefont {Nardelli}}, \bibinfo {author}
  {\bibfnamefont {M.}~\bibnamefont {Calandra}}, \bibinfo {author}
  {\bibfnamefont {R.}~\bibnamefont {Car}}, \bibinfo {author} {\bibfnamefont
  {C.}~\bibnamefont {Cavazzoni}}, \bibinfo {author} {\bibfnamefont
  {D.}~\bibnamefont {Ceresoli}}, \bibinfo {author} {\bibfnamefont
  {M.}~\bibnamefont {Cococcioni}}, \emph {et~al.},\ }\bibfield  {title}
  {\bibinfo {title} {Advanced capabilities for materials modelling with quantum
  espresso},\ }\href@noop {} {\bibfield  {journal} {\bibinfo  {journal}
  {Journal of physics: Condensed matter}\ }\textbf {\bibinfo {volume} {29}},\
  \bibinfo {pages} {465901} (\bibinfo {year} {2017})}\BibitemShut {NoStop}%
\bibitem [{\citenamefont {Sangalli}\ \emph {et~al.}(2019)\citenamefont
  {Sangalli}, \citenamefont {Ferretti}, \citenamefont {Miranda}, \citenamefont
  {Attaccalite}, \citenamefont {Marri}, \citenamefont {Cannuccia},
  \citenamefont {Melo}, \citenamefont {Marsili}, \citenamefont {Paleari},
  \citenamefont {Marrazzo} \emph {et~al.}}]{sangalli2019many}%
  \BibitemOpen
  \bibfield  {author} {\bibinfo {author} {\bibfnamefont {D.}~\bibnamefont
  {Sangalli}}, \bibinfo {author} {\bibfnamefont {A.}~\bibnamefont {Ferretti}},
  \bibinfo {author} {\bibfnamefont {H.}~\bibnamefont {Miranda}}, \bibinfo
  {author} {\bibfnamefont {C.}~\bibnamefont {Attaccalite}}, \bibinfo {author}
  {\bibfnamefont {I.}~\bibnamefont {Marri}}, \bibinfo {author} {\bibfnamefont
  {E.}~\bibnamefont {Cannuccia}}, \bibinfo {author} {\bibfnamefont
  {P.}~\bibnamefont {Melo}}, \bibinfo {author} {\bibfnamefont {M.}~\bibnamefont
  {Marsili}}, \bibinfo {author} {\bibfnamefont {F.}~\bibnamefont {Paleari}},
  \bibinfo {author} {\bibfnamefont {A.}~\bibnamefont {Marrazzo}}, \emph
  {et~al.},\ }\bibfield  {title} {\bibinfo {title} {Many-body perturbation
  theory calculations using the yambo code},\ }\href@noop {} {\bibfield
  {journal} {\bibinfo  {journal} {Journal of Physics: Condensed Matter}\
  }\textbf {\bibinfo {volume} {31}},\ \bibinfo {pages} {325902} (\bibinfo
  {year} {2019})}\BibitemShut {NoStop}%
\bibitem [{\citenamefont {Pizzi}\ \emph {et~al.}(2020)\citenamefont {Pizzi},
  \citenamefont {Vitale}, \citenamefont {Arita}, \citenamefont {Blügel},
  \citenamefont {Freimuth}, \citenamefont {G{\'{e}}ranton}, \citenamefont
  {Gibertini}, \citenamefont {Gresch}, \citenamefont {Johnson}, \citenamefont
  {Koretsune}, \citenamefont {Iba{\~{n}}ez-Azpiroz}, \citenamefont {Lee},
  \citenamefont {Lihm}, \citenamefont {Marchand}, \citenamefont {Marrazzo},
  \citenamefont {Mokrousov}, \citenamefont {Mustafa}, \citenamefont {Nohara},
  \citenamefont {Nomura}, \citenamefont {Paulatto}, \citenamefont
  {Ponc{\'{e}}}, \citenamefont {Ponweiser}, \citenamefont {Qiao}, \citenamefont
  {Thöle}, \citenamefont {Tsirkin}, \citenamefont {Wierzbowska}, \citenamefont
  {Marzari}, \citenamefont {Vanderbilt}, \citenamefont {Souza}, \citenamefont
  {Mostofi},\ and\ \citenamefont {Yates}}]{Pizzi2020}%
  \BibitemOpen
  \bibfield  {author} {\bibinfo {author} {\bibfnamefont {G.}~\bibnamefont
  {Pizzi}}, \bibinfo {author} {\bibfnamefont {V.}~\bibnamefont {Vitale}},
  \bibinfo {author} {\bibfnamefont {R.}~\bibnamefont {Arita}}, \bibinfo
  {author} {\bibfnamefont {S.}~\bibnamefont {Blügel}}, \bibinfo {author}
  {\bibfnamefont {F.}~\bibnamefont {Freimuth}}, \bibinfo {author}
  {\bibfnamefont {G.}~\bibnamefont {G{\'{e}}ranton}}, \bibinfo {author}
  {\bibfnamefont {M.}~\bibnamefont {Gibertini}}, \bibinfo {author}
  {\bibfnamefont {D.}~\bibnamefont {Gresch}}, \bibinfo {author} {\bibfnamefont
  {C.}~\bibnamefont {Johnson}}, \bibinfo {author} {\bibfnamefont
  {T.}~\bibnamefont {Koretsune}}, \bibinfo {author} {\bibfnamefont
  {J.}~\bibnamefont {Iba{\~{n}}ez-Azpiroz}}, \bibinfo {author} {\bibfnamefont
  {H.}~\bibnamefont {Lee}}, \bibinfo {author} {\bibfnamefont {J.-M.}\
  \bibnamefont {Lihm}}, \bibinfo {author} {\bibfnamefont {D.}~\bibnamefont
  {Marchand}}, \bibinfo {author} {\bibfnamefont {A.}~\bibnamefont {Marrazzo}},
  \bibinfo {author} {\bibfnamefont {Y.}~\bibnamefont {Mokrousov}}, \bibinfo
  {author} {\bibfnamefont {J.~I.}\ \bibnamefont {Mustafa}}, \bibinfo {author}
  {\bibfnamefont {Y.}~\bibnamefont {Nohara}}, \bibinfo {author} {\bibfnamefont
  {Y.}~\bibnamefont {Nomura}}, \bibinfo {author} {\bibfnamefont
  {L.}~\bibnamefont {Paulatto}}, \bibinfo {author} {\bibfnamefont
  {S.}~\bibnamefont {Ponc{\'{e}}}}, \bibinfo {author} {\bibfnamefont
  {T.}~\bibnamefont {Ponweiser}}, \bibinfo {author} {\bibfnamefont
  {J.}~\bibnamefont {Qiao}}, \bibinfo {author} {\bibfnamefont {F.}~\bibnamefont
  {Thöle}}, \bibinfo {author} {\bibfnamefont {S.~S.}\ \bibnamefont {Tsirkin}},
  \bibinfo {author} {\bibfnamefont {M.}~\bibnamefont {Wierzbowska}}, \bibinfo
  {author} {\bibfnamefont {N.}~\bibnamefont {Marzari}}, \bibinfo {author}
  {\bibfnamefont {D.}~\bibnamefont {Vanderbilt}}, \bibinfo {author}
  {\bibfnamefont {I.}~\bibnamefont {Souza}}, \bibinfo {author} {\bibfnamefont
  {A.~A.}\ \bibnamefont {Mostofi}},\ and\ \bibinfo {author} {\bibfnamefont
  {J.~R.}\ \bibnamefont {Yates}},\ }\bibfield  {title} {\bibinfo {title}
  {Wannier90 as a community code: new features and applications},\ }\href
  {https://doi.org/10.1088/1361-648x/ab51ff} {\bibfield  {journal} {\bibinfo
  {journal} {Journal of Physics: Condensed Matter}\ }\textbf {\bibinfo {volume}
  {32}},\ \bibinfo {pages} {165902} (\bibinfo {year} {2020})}\BibitemShut
  {NoStop}%
\bibitem [{\citenamefont {Ghosh}\ \emph {et~al.}(2019)\citenamefont {Ghosh},
  \citenamefont {Mondal}, \citenamefont {Kuo}, \citenamefont {Lue},
  \citenamefont {Nayak}, \citenamefont {Fujii}, \citenamefont {Vobornik},
  \citenamefont {Politano},\ and\ \citenamefont
  {Agarwal}}]{ghosh2019observation}%
  \BibitemOpen
  \bibfield  {author} {\bibinfo {author} {\bibfnamefont {B.}~\bibnamefont
  {Ghosh}}, \bibinfo {author} {\bibfnamefont {D.}~\bibnamefont {Mondal}},
  \bibinfo {author} {\bibfnamefont {C.-N.}\ \bibnamefont {Kuo}}, \bibinfo
  {author} {\bibfnamefont {C.~S.}\ \bibnamefont {Lue}}, \bibinfo {author}
  {\bibfnamefont {J.}~\bibnamefont {Nayak}}, \bibinfo {author} {\bibfnamefont
  {J.}~\bibnamefont {Fujii}}, \bibinfo {author} {\bibfnamefont
  {I.}~\bibnamefont {Vobornik}}, \bibinfo {author} {\bibfnamefont
  {A.}~\bibnamefont {Politano}},\ and\ \bibinfo {author} {\bibfnamefont
  {A.}~\bibnamefont {Agarwal}},\ }\bibfield  {title} {\bibinfo {title}
  {Observation of bulk states and spin-polarized topological surface states in
  transition metal dichalcogenide {Dirac} semimetal candidate {NiTe$_2$}},\
  }\href@noop {} {\bibfield  {journal} {\bibinfo  {journal} {Physical Review
  B}\ }\textbf {\bibinfo {volume} {100}},\ \bibinfo {pages} {195134} (\bibinfo
  {year} {2019})}\BibitemShut {NoStop}%
\bibitem [{\citenamefont {Mukherjee}\ \emph {et~al.}(2020)\citenamefont
  {Mukherjee}, \citenamefont {Jung}, \citenamefont {Weber}, \citenamefont {Xu},
  \citenamefont {Qian}, \citenamefont {Xu}, \citenamefont {Biswas},
  \citenamefont {Kim}, \citenamefont {Chapon}, \citenamefont {Watson} \emph
  {et~al.}}]{mukherjee2020fermi}%
  \BibitemOpen
  \bibfield  {author} {\bibinfo {author} {\bibfnamefont {S.}~\bibnamefont
  {Mukherjee}}, \bibinfo {author} {\bibfnamefont {S.~W.}\ \bibnamefont {Jung}},
  \bibinfo {author} {\bibfnamefont {S.~F.}\ \bibnamefont {Weber}}, \bibinfo
  {author} {\bibfnamefont {C.}~\bibnamefont {Xu}}, \bibinfo {author}
  {\bibfnamefont {D.}~\bibnamefont {Qian}}, \bibinfo {author} {\bibfnamefont
  {X.}~\bibnamefont {Xu}}, \bibinfo {author} {\bibfnamefont {P.~K.}\
  \bibnamefont {Biswas}}, \bibinfo {author} {\bibfnamefont {T.~K.}\
  \bibnamefont {Kim}}, \bibinfo {author} {\bibfnamefont {L.~C.}\ \bibnamefont
  {Chapon}}, \bibinfo {author} {\bibfnamefont {M.~D.}\ \bibnamefont {Watson}},
  \emph {et~al.},\ }\bibfield  {title} {\bibinfo {title} {Fermi-crossing
  type-{II} {Dirac} fermions and topological surface states in {NiTe$_2$}},\
  }\href@noop {} {\bibfield  {journal} {\bibinfo  {journal} {Scientific
  reports}\ }\textbf {\bibinfo {volume} {10}},\ \bibinfo {pages} {1} (\bibinfo
  {year} {2020})}\BibitemShut {NoStop}%
\bibitem [{\citenamefont {Nurmamat}\ \emph {et~al.}(2021)\citenamefont
  {Nurmamat}, \citenamefont {Eremeev}, \citenamefont {Wang}, \citenamefont
  {Yoshikawa}, \citenamefont {Kono}, \citenamefont {Kakoki}, \citenamefont
  {Muro}, \citenamefont {Jiang}, \citenamefont {Sun}, \citenamefont {Ye} \emph
  {et~al.}}]{nurmamat2021bulk}%
  \BibitemOpen
  \bibfield  {author} {\bibinfo {author} {\bibfnamefont {M.}~\bibnamefont
  {Nurmamat}}, \bibinfo {author} {\bibfnamefont {S.~V.}\ \bibnamefont
  {Eremeev}}, \bibinfo {author} {\bibfnamefont {X.}~\bibnamefont {Wang}},
  \bibinfo {author} {\bibfnamefont {T.}~\bibnamefont {Yoshikawa}}, \bibinfo
  {author} {\bibfnamefont {T.}~\bibnamefont {Kono}}, \bibinfo {author}
  {\bibfnamefont {M.}~\bibnamefont {Kakoki}}, \bibinfo {author} {\bibfnamefont
  {T.}~\bibnamefont {Muro}}, \bibinfo {author} {\bibfnamefont {Q.}~\bibnamefont
  {Jiang}}, \bibinfo {author} {\bibfnamefont {Z.}~\bibnamefont {Sun}}, \bibinfo
  {author} {\bibfnamefont {M.}~\bibnamefont {Ye}}, \emph {et~al.},\ }\bibfield
  {title} {\bibinfo {title} {Bulk {Dirac} cone and highly anisotropic
  electronic structure of {NiTe$_2$}},\ }\href@noop {} {\bibfield  {journal}
  {\bibinfo  {journal} {Physical Review B}\ }\textbf {\bibinfo {volume}
  {104}},\ \bibinfo {pages} {155133} (\bibinfo {year} {2021})}\BibitemShut
  {NoStop}%
\bibitem [{\citenamefont {Clark}\ \emph {et~al.}(2019)\citenamefont {Clark},
  \citenamefont {Mazzola}, \citenamefont {Markovi{\'c}}, \citenamefont {Riley},
  \citenamefont {Feng}, \citenamefont {Yang}, \citenamefont {Sumida},
  \citenamefont {Okuda}, \citenamefont {Fujii}, \citenamefont {Vobornik} \emph
  {et~al.}}]{clark2019general}%
  \BibitemOpen
  \bibfield  {author} {\bibinfo {author} {\bibfnamefont {O.~J.}\ \bibnamefont
  {Clark}}, \bibinfo {author} {\bibfnamefont {F.}~\bibnamefont {Mazzola}},
  \bibinfo {author} {\bibfnamefont {I.}~\bibnamefont {Markovi{\'c}}}, \bibinfo
  {author} {\bibfnamefont {J.~M.}\ \bibnamefont {Riley}}, \bibinfo {author}
  {\bibfnamefont {J.}~\bibnamefont {Feng}}, \bibinfo {author} {\bibfnamefont
  {B.}~\bibnamefont {Yang}}, \bibinfo {author} {\bibfnamefont {K.}~\bibnamefont
  {Sumida}}, \bibinfo {author} {\bibfnamefont {T.}~\bibnamefont {Okuda}},
  \bibinfo {author} {\bibfnamefont {J.}~\bibnamefont {Fujii}}, \bibinfo
  {author} {\bibfnamefont {I.}~\bibnamefont {Vobornik}}, \emph {et~al.},\
  }\bibfield  {title} {\bibinfo {title} {A general route to form
  topologically-protected surface and bulk {Dirac} fermions along high-symmetry
  lines},\ }\href@noop {} {\bibfield  {journal} {\bibinfo  {journal}
  {Electronic Structure}\ }\textbf {\bibinfo {volume} {1}},\ \bibinfo {pages}
  {014002} (\bibinfo {year} {2019})}\BibitemShut {NoStop}%
\bibitem [{\citenamefont {Karn}\ and\ \citenamefont {Awana}(2023)}]{KARN2023}%
  \BibitemOpen
  \bibfield  {author} {\bibinfo {author} {\bibfnamefont {N.}~\bibnamefont
  {Karn}}\ and\ \bibinfo {author} {\bibfnamefont {V.}~\bibnamefont {Awana}},\
  }\bibfield  {title} {\bibinfo {title} {Band topology and non-trivial surface
  states in type-ii dirac semi-metal x(ni, pd)te$_2$},\ }\bibfield  {journal}
  {\bibinfo  {journal} {Materials Today: Proceedings}\ }\href
  {https://doi.org/https://doi.org/10.1016/j.matpr.2023.05.202}
  {https://doi.org/10.1016/j.matpr.2023.05.202} (\bibinfo {year}
  {2023})\BibitemShut {NoStop}%
\bibitem [{\citenamefont {Monteiro}\ \emph {et~al.}(2017)\citenamefont
  {Monteiro}, \citenamefont {Marciniak}, \citenamefont {Jurelo}, \citenamefont
  {Siqueira}, \citenamefont {Dias},\ and\ \citenamefont
  {J{\'u}nior}}]{monteiro2017synthesis}%
  \BibitemOpen
  \bibfield  {author} {\bibinfo {author} {\bibfnamefont {J.~F. H.~L.}\
  \bibnamefont {Monteiro}}, \bibinfo {author} {\bibfnamefont {M.~B.}\
  \bibnamefont {Marciniak}}, \bibinfo {author} {\bibfnamefont {A.~R.}\
  \bibnamefont {Jurelo}}, \bibinfo {author} {\bibfnamefont {E.~C.}\
  \bibnamefont {Siqueira}}, \bibinfo {author} {\bibfnamefont {F.~T.}\
  \bibnamefont {Dias}},\ and\ \bibinfo {author} {\bibfnamefont {J.~L.~P.}\
  \bibnamefont {J{\'u}nior}},\ }\bibfield  {title} {\bibinfo {title} {Synthesis
  and microstructure of {NiTe$_2$}},\ }\href@noop {} {\bibfield  {journal}
  {\bibinfo  {journal} {Journal of Crystal Growth}\ }\textbf {\bibinfo {volume}
  {478}},\ \bibinfo {pages} {129} (\bibinfo {year} {2017})}\BibitemShut
  {NoStop}%
\bibitem [{\citenamefont {Ferreira}\ \emph {et~al.}(2021)\citenamefont
  {Ferreira}, \citenamefont {Manesco}, \citenamefont {Dorini}, \citenamefont
  {Correa}, \citenamefont {Weber}, \citenamefont {Machado},\ and\ \citenamefont
  {Eleno}}]{ferreira2021strain}%
  \BibitemOpen
  \bibfield  {author} {\bibinfo {author} {\bibfnamefont {P.~P.}\ \bibnamefont
  {Ferreira}}, \bibinfo {author} {\bibfnamefont {A.~L.}\ \bibnamefont
  {Manesco}}, \bibinfo {author} {\bibfnamefont {T.~T.}\ \bibnamefont {Dorini}},
  \bibinfo {author} {\bibfnamefont {L.~E.}\ \bibnamefont {Correa}}, \bibinfo
  {author} {\bibfnamefont {G.}~\bibnamefont {Weber}}, \bibinfo {author}
  {\bibfnamefont {A.~J.}\ \bibnamefont {Machado}},\ and\ \bibinfo {author}
  {\bibfnamefont {L.~T.}\ \bibnamefont {Eleno}},\ }\bibfield  {title} {\bibinfo
  {title} {Strain engineering the topological type-{II} {Dirac} semimetal
  {NiTe$_2$}},\ }\href@noop {} {\bibfield  {journal} {\bibinfo  {journal}
  {Physical Review B}\ }\textbf {\bibinfo {volume} {103}},\ \bibinfo {pages}
  {125134} (\bibinfo {year} {2021})}\BibitemShut {NoStop}%
\bibitem [{\citenamefont {Elias}\ \emph {et~al.}(2011)\citenamefont {Elias},
  \citenamefont {Gorbachev}, \citenamefont {Mayorov}, \citenamefont {Morozov},
  \citenamefont {Zhukov}, \citenamefont {Blake}, \citenamefont {Ponomarenko},
  \citenamefont {Grigorieva}, \citenamefont {Novoselov}, \citenamefont
  {Guinea},\ and\ \citenamefont {Geim}}]{Elias2011}%
  \BibitemOpen
  \bibfield  {author} {\bibinfo {author} {\bibfnamefont {D.~C.}\ \bibnamefont
  {Elias}}, \bibinfo {author} {\bibfnamefont {R.~V.}\ \bibnamefont
  {Gorbachev}}, \bibinfo {author} {\bibfnamefont {A.~S.}\ \bibnamefont
  {Mayorov}}, \bibinfo {author} {\bibfnamefont {S.~V.}\ \bibnamefont
  {Morozov}}, \bibinfo {author} {\bibfnamefont {A.~A.}\ \bibnamefont {Zhukov}},
  \bibinfo {author} {\bibfnamefont {P.}~\bibnamefont {Blake}}, \bibinfo
  {author} {\bibfnamefont {L.~A.}\ \bibnamefont {Ponomarenko}}, \bibinfo
  {author} {\bibfnamefont {I.~V.}\ \bibnamefont {Grigorieva}}, \bibinfo
  {author} {\bibfnamefont {K.~S.}\ \bibnamefont {Novoselov}}, \bibinfo {author}
  {\bibfnamefont {F.}~\bibnamefont {Guinea}},\ and\ \bibinfo {author}
  {\bibfnamefont {A.~K.}\ \bibnamefont {Geim}},\ }\bibfield  {title} {\bibinfo
  {title} {Dirac cones reshaped by interaction effects in suspended graphene},\
  }\href {https://doi.org/10.1038/nphys2049} {\bibfield  {journal} {\bibinfo
  {journal} {Nature Physics}\ }\textbf {\bibinfo {volume} {7}},\ \bibinfo
  {pages} {701–704} (\bibinfo {year} {2011})}\BibitemShut {NoStop}%
\bibitem [{\citenamefont {Wu}\ \emph {et~al.}(2018)\citenamefont {Wu},
  \citenamefont {Zhang}, \citenamefont {Song}, \citenamefont {Troyer},\ and\
  \citenamefont {Soluyanov}}]{WU2017}%
  \BibitemOpen
  \bibfield  {author} {\bibinfo {author} {\bibfnamefont {Q.}~\bibnamefont
  {Wu}}, \bibinfo {author} {\bibfnamefont {S.}~\bibnamefont {Zhang}}, \bibinfo
  {author} {\bibfnamefont {H.-F.}\ \bibnamefont {Song}}, \bibinfo {author}
  {\bibfnamefont {M.}~\bibnamefont {Troyer}},\ and\ \bibinfo {author}
  {\bibfnamefont {A.~A.}\ \bibnamefont {Soluyanov}},\ }\bibfield  {title}
  {\bibinfo {title} {Wanniertools : An open-source software package for novel
  topological materials},\ }\href
  {https://doi.org/https://doi.org/10.1016/j.cpc.2017.09.033} {\bibfield
  {journal} {\bibinfo  {journal} {Computer Physics Communications}\ }\textbf
  {\bibinfo {volume} {224}},\ \bibinfo {pages} {405 } (\bibinfo {year}
  {2018})}\BibitemShut {NoStop}%
\end{thebibliography}%






\end{document}